\documentclass[final,5p,times,twocolumn]{elsarticle}

\usepackage{lineno,hyperref}
\modulolinenumbers[5]

\journal{Elsevier}
\usepackage{graphicx}
\usepackage[table,xcdraw]{xcolor}
\usepackage[ansinew]{inputenc}
\usepackage{array}
\usepackage{makecell}
\usepackage{color}
\usepackage{xcolor}
\usepackage{amsmath}
\usepackage{amsmath,amssymb,amsfonts}
\usepackage{algorithmic}
\usepackage{graphicx}
\usepackage{caption}
\usepackage{textcomp}
\usepackage{longtable}
\usepackage{enumitem}
\usepackage{rotating}
\usepackage{subfigure}
\usepackage{hyperref}
\usepackage{amsxtra}
\usepackage{natbib}
\usepackage{amssymb}
\usepackage{subfigure}
\usepackage{latexsym}
\usepackage{caption}
\usepackage{amsmath,amsfonts,amssymb,amsthm}
\usepackage[ruled,vlined]{algorithm2e}
\usepackage{multirow}
\definecolor{Gray}{gray}{0.9}
\usepackage{balance}
\usepackage{xcolor}
\usepackage{amsmath}
\usepackage{tikz}
\usetikzlibrary{shapes}
\usepackage{color,soul}
\usepackage{comment}

\PassOptionsToPackage{hyphens}{url}\usepackage{hyperref}

\begin{document}
\begin{frontmatter}

\title{Open RAN Security: Challenges and Opportunities}


\author[1]{Madhusanka~Liyanage}
\ead{madhusanka@ucd.ie}
\author[2]{An Braeken}
\ead{an.braeken@vub.be}
\author[3] {Shahriar~Shahabuddin}
\ead{shahriar.shahabuddin@nokia.com}
\author[4] {Pasika Ranaweera}
\ead{pasika.ranaweera@ucdconnect.ie}

\address[1]{School of Computer Science, University Collage Dublin, Ireland and Centre for Wireless Communications, University of Oulu, Finland}

\address[2]{Department of engineering, technology (INDI), Vrije Universiteit Brussel, Belgium}

\address[3]{Mobile Networks, Nokia, Dallas, TX, USA, and Centre for Wireless Communications, University of Oulu, Finland}

\address[4]{School of Computer Science, University Collage Dublin, Ireland}

\begin{abstract}
Open RAN (ORAN, O-RAN) represents a novel industry-level standard for RAN (Radio Access Network), which defines interfaces that support inter-operation between vendors' equipment and offer network flexibility at a lower cost. Open RAN integrates the benefits and advancements of network softwarization and Artificial Intelligence to enhance the operation of RAN devices and operations. Open RAN offers new possibilities so that different stakeholders can develop the RAN solution in this open ecosystem. However, the benefits of Open RAN bring new security and privacy challenges. As Open RAN offers an entirely different RAN configuration than what exists today, it could lead to severe security and privacy issues if mismanaged, and stakeholders are understandably taking a cautious approach towards the security of Open RAN deployment. In particular, this paper provides a deep analysis of the security and privacy risks and challenges associated with Open RAN architecture. Then, it discusses possible security and privacy solutions to secure Open RAN architecture and presents relevant security standardization efforts relevant to Open RAN security. Finally, we discuss how Open RAN can be used to deploy more advanced security and privacy solutions in 5G and beyond RAN. 
\end{abstract}
\medskip
\end{frontmatter}

\section{Introduction}
\label{Sec:Introduction}


\begin{table*}[h!]
\centering
\footnotesize
\caption{Summary of Important Acronyms}
\label{table:acronyms}
\begin{tabular}{|p{1.5cm} p{6.5cm}|p{1.5cm} p{6.5cm}|}
\hline
\rowcolor{gray!30}
\textbf{Acronym} & \textbf{Definition} & \textbf{Acronym} & \textbf{Definition}\\
\hline
\hline

3GPP  &	3rd Generation Partnership Project	&	
5G  &	Fifth Generation	\\	
AI  &	Artificial Intelligence			&
AICPA  &	American Institute of Certified Public Accountants	 	\\	
API  &	Application Programming Interface		&	
BBU	  & 	Baseband Unit			\\
CA	  & 	Certificate Authority			&
CIA	  & 	Confidentiality, Integrity, Availability		\\	
CICA	  & 	Canadian Institute of Chartered Accountants		&	
CI/CD	  & 	Continuous Integration / Continuous Delivery 	\\		
CN	  & 	Container			&
CNF	  & 	Containerized or Cloud-Native Network Function 	\\			
COTS	  & 	Commercial Off-The-Shelf			&
CPU	  & 	Central Processing Unit	 	\\		
(D)DoS	  & 	 (Distributed) Denial of Service	&		
DL	  & 	Downlink 	\\			
DU	  & 	Distributed Unit			&
E2E	  & 	End-to-end 	 	\\		
eCPRI  & 		Enhanced Common Public Radio Interface		&	
EI	  & 	Election Infrastructure	 	\\		
ETSI	  & 	European Telecommunications Standards Institute		&	
FH	  & 	Fronthaul	 	\\		
FRANS	  & 	Fair, Reasonable and Non Discriminatory		&	
FTP	  & 	File Transfer Protocol	 	\\		
GDPR	  & 	General Data Protection Regulation			&
gNB	  & 	Next generation NodeB 	\\			
GPS	  & 	Global Positioning System			&
GUTI	  & 	Global Unique Temporary Identifier	 	\\		
HTTP	  & 	Hypertext Transfer Protocol			&
HW	  & 	Hardware		 	\\	
IoT	  & 	Internet of Things			&
IP	  & 	Intellectual Property 	\\			
IPSec  & 		Internet Protocol Security		&	
JTAG	  & 	Joint Test Action Group	 	\\		
LTE-M	  & 	Long Term Evolution for Machines	&		
MI	  & 	Model Inversion		 	\\	
MIMO	  & 	Multiple Input Multiple Output		&	
MITM	  & 	Man In The Middle		 	\\	
ML	  & 	Machine Learning			&
M-Plane  & 		Management Plane	 	\\		
MTBF	  & 	Mean Time Between Failures		&	
Near-RT  & 		Near-Real-Time		 	\\	
NF	  & 	Network Function			&
Non-RT  & 		Non-Real-Time	 	\\		
NVD	  & 	National Vulnerability Database		&	
OAM	  & 	Operations, Administration and Maintenance	 	\\		
O-Cloud  & 		Open Cloud			&
O-CU	  & 	Open Centralized Unit 	\\			
O-DU	  & 	Open Distributed Unit			&
OFH	  & 	Open Fronthaul		 	\\	
Open RAN	  & 	Open Radio Access Network		&	
O-RU	  & 	Open RAN Radio Unit	 	\\		
OS	  & 	Operating System			&
OSS	  & 	Operations Support Systems	 	\\		
PBCH	  & 	Physical Broadcast Channel			&
PDCCH  & 		Physical Downlink Control Channel	 	\\		
PNF	  & 	Physical Network Function			&
PTP	  & 	Precision Time Protocol		 	\\	
RAM	  & 	Random Access Memory			&
RAN	  & 	Radio Access Network		 	\\	
rApp	  & 	 non-real-time intelligence Application		&	
RIC	  & 	Radio Access Network Intelligent Controller	 	\\		
RRM	  & 	Radio Resource Management			&
RRU	  & 	Remote Radio Unit	 	\\		
RU	  & 	Radio Unit			&
SSH	  & 	Secure Shell		 	\\	
SI	  & 	System Integration			&
SMO	  & 	Service Management and Orchestration	 	\\		
S Plane	  & 	Synchronization Plane			&
SRM	  & 	Supplier Relationship Management 	\\			
SUPI	  & 	Subscription Permanent Identifier	&		
SW	  & 	Software		 	\\	
SQL	  & 	Structured Query Language		&	
TCP	  & 	Transmission Control Protocol	 	\\		
TPM	  & 	Trusted Platform Module			&
UDP	  & 	User Datagram Protocol	 	\\		
UE	  & 	User Equipment			&
UL	  & 	Uplink		 	\\	
U Plane, UP  & 		User plane			&
vCU	  & 	Virtual Computational Unit	 	\\		
VIP	  & 	Very Important Person			&
VM	  & 	Virtual Machine		 	\\	
VNF	  & 	Virtual Network Function		&	
vRAN	  & 	virtualized Radio Access Network 	\\			
xApp  & 	near-real-time intelligence Application		& & 	\\	

\hline
\end{tabular}
\end{table*}


Mobile network communications are becoming one of the critical enablers of the current digital economy and interconnecting national critical infrastructure-based services \cite{crit}. The number of mobile subscribers and different mobile-based services is increasing in a rapid phase all across the globe \cite{num123}. However, the radio spectrum is still a scarce resource, and optimal utilization of radio resources is critical for developing a telecommunication network \cite{faulhaber2003spectrum}. Thus, the orchestration and management of radio resources or the Radio Access Network (RAN) are also evolved with each mobile generation. The early mobile generation mobile networks architectures such as  2G and 3G had controllers responsible for orchestration and management of RAN and its resources \cite{gindraux20022g}. The flat network architecture in 4G enables a new interface (i.e., X2) to support base station level communication to handle RAN resource allocation \cite{dahlman20134g}. However, the RAN of existing mobile network generations is still based on monolithic building blocks. Thus, RAN functions of existing networks, including most of the 5G network, are still contained with the proprietary vendor-specific devices called Baseband Units (BBUs) at the base stations \cite{parvez2018survey}. However, this approach leads to the proverbial vendor lock-in RAN since different RAN vendors can design their flavor of RAN equipment. This has eliminated the possibility for MNOs (Mobile Network Operators) to get mix-and-match services from other RAN vendors.

The introduction of the network softwarization concept in 5G \cite{nguyen2021empowering,condoluci2018softwarization,yang2013openran} and added intelligence in beyond 5G networks have opened up a promising solution, called Open RAN, to mitigate this issue \cite{yang2013openran,gavrilovska2020cloud}. The Open RAN Alliance \cite{umesh2020ran} went back to the controller concept to enable best-of-breed Open RAN. Open Radio Access Networks (Open RANs), also known as ORANs or O-RANs, have been considered one of the most exciting RAN concepts, designed for 5G and beyond wireless systems. Open RAN promotes openness and added intelligence for RAN network elements that could overcome the limitations of existing RAN technologies, \cite{25,bonati2022openran}. The feature of openness allows smaller and new players in the RAN market to deploy their customized services, while the feature of intelligence is to increase automation and performance by optimizing the RAN elements and network resources. Moreover, Open RAN offers many RAN solutions and elements to the network operators to be more open and flexible. Further, the network operators can shorten the time-to-market of new applications and services to maximize the overall revenue because of the virtualization feature. Thus, the added intelligence in Open RAN could offer superior benefits even to existing network softwarization based virtual RAN and cloud RAN concepts.

There are two major Open RAN organizations, i.e., Telecom Infra Project (TIP)\cite{TIP} and the O-RAN alliance\cite{ORAN} who are working on the advancement of Open RAN realization. The TIP's OpenRAN program is an initiative that focuses on developing solutions for future RANs based on disaggregation of multi-vendor hardware, open interfaces, and software. O-RAN alliance is another Open RAN organization that mainly focuses on defining and enforcing new standards for Open RAN to ensure interoperability among the different vendors. At the beginning of 2020, a liaison agreement between TIP and O-RAN was made to ensure their alignment in developing interoperable Open RAN solutions. OpenRAN development of TIP has similar original goal similar to O-RAN. Thus, we use the term ``Open RAN'' throughout the paper with refer to both OpenRAN and O-RAN development efforts.

O-RAN refers to the O-RAN Alliance or designated specification.

However, the benefits of Open RAN come at challenges, e.g., security, deterministic latency, and real-time control \cite{16, gavrilovska2020cloud, polese2022understanding}. Among these factors, the security in Open RAN is quite essential. As 5G and beyond networks are responsible for interconnecting many Internet protocol Telephony (IpT) based critical national infrastructure, attacks on future telecommunication networks will have a ripple effect \cite{mantas2015security,batalla2020security}. Some devastating examples caused by such attacks are smart cities and factories shutting down, a complete black-out of the power grids and water supplies, a fall-out of the transportation infrastructure with crashes by autonomous vehicles, etc. \cite{soldani20195g, ahmad2018overview}. All these challenges demand significant effort from the research and industry communities to standardize and implement security for all the sections of 5G and beyond networks, including  Open RAN networks \cite{tataria20216g}. Specially, the decentralization of control functions with Open RAN increases the number of threat vectors and the surface area for attacks.

Open RAN has distinct features that bring intelligence to future networks. While AI helps overcome various challenges of 6G Open RANs via intelligent and data-driven solutions, it can hurt the security of RAN. Attackers can target the AI systems or even use AI-based attacks to jeopardize the operation of the Open RAN system. Thus, Open RAN will now be vulnerable to AI-related attacks such as denial-of-service (DoS) \cite{needham1993denial}, spoofing \cite{van2018classification} and malicious data injection \cite{illiano2015detecting} could affect the AI. For instance, AI training can be manipulated in an Open RAN spectrum access system by inserting fake signals. In addition, the integration of network softwarization will add a whole new set of security attacks related to virtualization. Similar to the 5G core and edge networks, now Open RAN needs to tackle softwarization associated attacks such as Virtual Network Function)/Cloud Network Function (VNF/CNF) manipulation \cite{kawashima2021vision}, Virtual Machine (VM) misconfiguration \cite{jarraya2015verification}, log leak attacks \cite{wright2003defending}. In addition, open interfaces defined in Open  RAN will introduce another set of security and privacy vulnerability. Thus, it is necessary to develop correct security and privacy solutions to mitigate these new Open RAN-related security and privacy solutions at the radio network level. Existing security mechanisms, frameworks, and governance approaches will need to be upgraded to operate in open multi-vendor controlled Ecosystem.

 \begin{table*}
\caption{Summary of publications relevant to Open RAN security}
\label{tab:related_work}
\renewcommand{\arraystretch}{1}
  \begin{tabular}{| p{1.5cm}|p{0.33cm}|p{0.33cm}|p{0.33cm}|p{0.33cm}|p{0.33cm}|p{11.5cm}|}
  \hline
      \rowcolor{Gray}
    	\multicolumn{1}{|c|}{\textbf{Year \& Ref.}} 
         & \multicolumn{1}{|c|}{\textbf{{\rotatebox[origin=c]{90}{Open RAN Architecture}}}}
         &\multicolumn{1}{|c|}{ \textbf{{\rotatebox[origin=c]{90}{Open RAN Security Flaws}}}}
         &\multicolumn{1}{|c|}{\textbf{{\rotatebox[origin=c]{90}{Open RAN Security Solutions}}}}
         &\multicolumn{1}{|c|}{\textbf{{\rotatebox[origin=c]{90}{Open RAN Security Benefits}}}}
         &\multicolumn{1}{|c|}{\textbf{{\rotatebox[origin=c]{90}{Research Directions }}}}
         &\multicolumn{1}{|c|}{\textbf{Remarks}}\\ [10ex]
    \hline
    \hline
        \multicolumn{1}{|c|}{2022~\cite{wypior2022open}} & \cellcolor{green!15} H & \cellcolor{red!15} L & \cellcolor{red!15} L & \cellcolor{red!15} L &  \cellcolor{red!15} L & A review article on RAN evolution towards open models and potential Open RAN benefits and market trends\\ [2ex]
    \hline
        \multicolumn{1}{|c|}{2022~\cite{7}} & \cellcolor{yellow!20} M & \cellcolor{red!15} L & \cellcolor{red!15} L & \cellcolor{red!15} L &  \cellcolor{red!15} L & \textcolor{black}{This article discusses Open RAN deployment with a focus on 5G network device security}\\ [2ex]
    \hline
        \multicolumn{1}{|c|}{2021~\cite{13}} & \cellcolor{green!15} H & \cellcolor{yellow!20} M & \cellcolor{green!15} H & \cellcolor{red!15} L &  \cellcolor{red!15} L & A whitepaper by Altiostar on the security of Open RAN which presents a method to implement Open RAN with a zero-trust security framework\\ [2ex]
    \hline
        \multicolumn{1}{|c|}{2021~\cite{garcia2021ran}} & \cellcolor{green!15} H & \cellcolor{red!15} L & \cellcolor{red!15} L & \cellcolor{red!15} L &  \cellcolor{red!15} L & \textcolor{black}{An article that summarizes Open RAN specifications focusing on proposed architecture and building blocks}\\ [2ex] 
         \hline
    \multicolumn{1}{|c|}{2021~\cite{dryjanski2021toward}} & \cellcolor{green!15} H & \cellcolor{red!15} L & \cellcolor{red!15} L & \cellcolor{red!15} L &  \cellcolor{red!15} L & \textcolor{black}{This article presents an analysis of an Open RAN system with the aid of a traffic steering use case implemented in a modular way}\\ [2ex]    
    \hline
        \multicolumn{1}{|c|}{2021~\cite{4}} & \cellcolor{green!15} H & \cellcolor{green!15} H & \cellcolor{red!15} L & \cellcolor{red!15} L &  \cellcolor{red!15} L & A technical specification by O-RAN alliance on Open RAN security threat modeling and remediation analysis\\ [2ex]
    \hline
        \multicolumn{1}{|c|}{2021~\cite{abdalla2021toward}} & \cellcolor{green!15} H & \cellcolor{green!15} H & \cellcolor{red!15} L & \cellcolor{yellow!20} M &  \cellcolor{red!15} L & A pre-print which identifies the limitations of current Open RAN architecture and the technologies and opportunities for research and development to overcome them\\ [2ex]
    \hline   
        \multicolumn{1}{|c|}{2021~\cite{ORANSpec}} & \cellcolor{yellow!20} M & \cellcolor{yellow!20} M & \cellcolor{green!15} H & \cellcolor{red!15} L &  \cellcolor{red!15} L & A technical specification by O-RAN alliance on the security requirements and security controls per Open RAN defined interface and Open RAN defined network function\\ [2ex]
    \hline  
        \multicolumn{1}{|c|}{2021~\cite{transport_sec_ORAN}} & \cellcolor{yellow!20} M & \cellcolor{red!15} L & \cellcolor{yellow!20} M & \cellcolor{red!15} L &  \cellcolor{red!15} L & Presented an analysis to demonstrate the urgent need to protect Open RAN fronthaul and proposed a security protocol as a potential solution\\ [2ex]
    \hline  
        \multicolumn{1}{|c|}{2020~\cite{5}} & \cellcolor{red!15} L & \cellcolor{yellow!20} M & \cellcolor{red!15} L & \cellcolor{red!15} L &  \cellcolor{red!15} L & A whitepaper by Ericsson on Open RAN security considerations that ensure an open and interoperable RAN is secure by design\\ [2ex]
    \hline
        \multicolumn{1}{|c|}{2020~\cite{gavrilovska2020cloud}} & \cellcolor{green!15} H & \cellcolor{red!15} L & \cellcolor{red!15} L & \cellcolor{red!15} L &  \cellcolor{red!15} L & This article presents the basic functionalities and current research trends on C-RAN and its derivatives such as vRAN and Open RAN\\ [2ex]
    \hline       
        \multicolumn{1}{|c|}{2017~\cite{CRAN_security}} & \cellcolor{red!15} L & \cellcolor{yellow!20} M & \cellcolor{yellow!20} M & \cellcolor{red!15} L &  \cellcolor{red!15} L & A survey of C-RAN security flaws and solutions where many threats and solutions are relevant for Open RAN.\\ [2ex]
    \hline  
         \multicolumn{1}{|c|}{This Paper} & \cellcolor{green!15} H & \cellcolor{green!20} H & \cellcolor{green!15} H & \cellcolor{green!20} H &  \cellcolor{green!15} H & A comprehensive security analysis of Open RAN which througly discusses the Open RAN security architecture, security flaws and solutions and security benefits of Open RAN\\ [2ex]
    \hline   
  \end{tabular}

\begin{flushleft}
\begin{center}

\begin{tikzpicture}

\node (rect) at (6,2) [draw,thick,minimum width=1cm,minimum height=0.4cm, fill= red!15, label=0:Low Coverage: Did not Consider the factor or only very briefly discussed it through mentioning it in passing] {L};
\node (rect) at (6,2.5) [draw,thick,minimum width=1cm,minimum height=0.4cm, fill= yellow!20, label=0:Medium Coverage: Partially considers the factor (leaves out vital aspects or discusses it in relation to other factors)] {M};
\node (rect) at (6,3) [draw,thick,minimum width=1cm,minimum height=0.4cm, fill= green!15, label=0:High Coverage: Consider the factor in reasonable or high detail] {H};
\end{tikzpicture}


\end{center}

\end{flushleft}
  
\end{table*}

On the other hand, added features of Open RAN can bring security and privacy advantages over traditional RAN. Open RAN can also build upon the security enhancements already enabled by 5G and allow the operator to control the network's security entirely, ultimately enhancing the operational security of their network. Less hardware dependency and support for complete software control in Open RAN allow isolating security breaches quickly and intelligently, reducing the impact of security risk. In addition, these features reduce the risks associated with security mechanism upgrades. Moreover, the modularity supported by the open interface in Open RAN allows the security and privacy deployments to support continuous integration/continuous delivery (CI/CD) operating model \cite{bobrovskis2018survey}. The CI/CD model supports seamless and effective security management against the security vulnerability in Open RAN.

Moreover,  Open RAN enables the possibility for zero-touch and frequent software updates \cite{dutta2021challenge}, which is more transparent, fast,  secure, and low cost than the software upgrades in a traditional network. Finally, standardization of open interfaces can also reduce the security risks to a certain extent as it can help detect incongruencies and offer concrete steps to monitor the network. Thus, it is crucial to identify these new security benefits and rectify them correctly in future RAN deployments.

\subsection{Motivation}

The research on Open RAN security is still in its infancy. As Open RAN advocates open interfaces, it is imperative to analyze the security vulnerabilities and their mitigation of Open RAN in parallel to their system architecture development. The reason is that without a proper security framework in place, the idea of an open network might not be an attractive solution to the network operators. This is especially true in this era of complex geopolitics, where global powers are increasingly concerned about wireless infrastructure security. Table~\ref{tab:related_work} summarizes existing research works about Open RAN security. The table highlights the lack of a comprehensive Open RAN security analysis in the literature. Most existing Open RAN-related publications focused on architecture, interfaces, and algorithms, while security was a secondary topic. 
A couple of technical specifications from Open RAN alliance present the security flaws and solutions of Open RAN in~\cite{4}, and \cite{ORANSpec}, respectively. However, they are not comprehensive because they lack a thorough discussion either on the solutions or the flaws. They also do not present the Open RAN security benefits and discuss any research directions. Similarly, other publications presented in Table~\ref{tab:related_work} fail to provide a comprehensive analysis of Open RAN security.

\color{black}
\subsection{Our Contribution}


To the best of the authors' knowledge, this is the first attempt to provide a comprehensive security analysis of Open RAN. The main contributions of this article are presented below.

\begin{itemize}
    \item \textbf{Classification of security-related risks}:  A taxonomy distinguishing the risks present in Open RAN, is provided. Each of these risks is elaborated concerning a description of impact. 
    
    \item \textbf{Present Open RAN specific security solutions}: 
    Unique solutions for Open RAN security vulnerabilities based on blockchain, physical layer, and AI have been presented.
    
    \item \textbf{Overview of general mistakes, consequences, and mitigation}: A summary of the general design errors pertaining to Open RAN, their consequences, and potential mitigation are presented.
    
    \item \textbf{Discussion on Open RAN security benefits}: A list of security benefits specific to Open RAN, and already available in V-RAN and 5G networks are presented. 
     
\end{itemize}

\subsection{Outline}

The rest of the paper is organized as follows. Section~\ref{sec:OpenRAN} presents the overview of Open RAN architecture and the difference from the conventional RAN architectures. The threat vectors and security risks associated with Open RAN are presented in Section~\ref{sec:threats}. Several solutions for the security threats and vulnerabilities of Open RAN are elaborated in Section~\ref{sec:solution}. Section~\ref{sec:Benefits} presents the security benefits of Open RAN implementation. Discussion and lessons learned towards realizing an Open RAN architecture are portrayed in Section~\ref{sec:discussion}. Finally, Section~\ref{Sec:Conclusion} concludes the paper.

\newpage
\section{Brief Overview of Open RAN Architecture} 
\label{sec:OpenRAN}

\begin{figure}[h]
\centering
\includegraphics[width=0.5\textwidth]{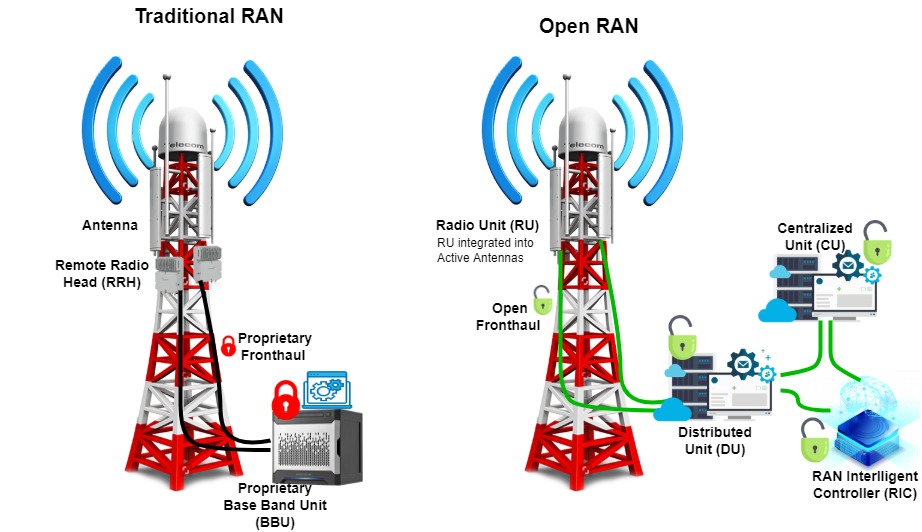}
\caption{High level Comparison of Open RAN with Traditional RAN} 
\label{Fig:ORANvsTRAN}
\end{figure}

Unlike traditional RAN technology, Open RAN decouples hardware and software bonds in proprietary RAN equipment. This feature offers more flexibility for mobile operators to deploy and upgrade their RAN segment \cite{yang2013openran, 25, johnson2022nexran}. Figure \ref{Fig:ORANvsTRAN} illustrated the key differences of traditional and Open RAN architectures. 

The Open RAN architecture is proposed to enable three main goals \cite{garcia2021ran, gavrilovska2020cloud}, i.e.;

\begin{itemize}
\item Cloudification: The goal is to support cloud-native RAN functions via disaggregated hardware and software components.
\item Intelligence and automation: The goal is to utilize advanced AI/ML capabilities to enable automated management and orchestration in RAN
\item Open internal RAN interfaces:  The goal is to support various Open RAN interfaces, including interfaces defined by 3GPP.
\end{itemize}

As illustrated in Figure \ref{Fig:ORANvsTRAN},  the RAN in Open RAN architecture is disaggregated into four main building blocks, i.e., the Radio Unit (RU), the Distributed Unit (DU), the Centralised Unit (CU), and RAN Intelligence Controller (RIC). The RU is located with antennas, and it is responsible for transmitting, receiving, amplifying, and digitizing the radio frequency signals. The former BBU (Based Band Unit) is now disaggregated into DU and CU. They are the computation parts of the base station. Here, DU is physically located closer to RU, while CU can be located closer to the Core. RIC is possible for taking the intelligent and automated decisions related to RAN.

\begin{figure}[h]
\centering
\includegraphics[width=0.5\textwidth]{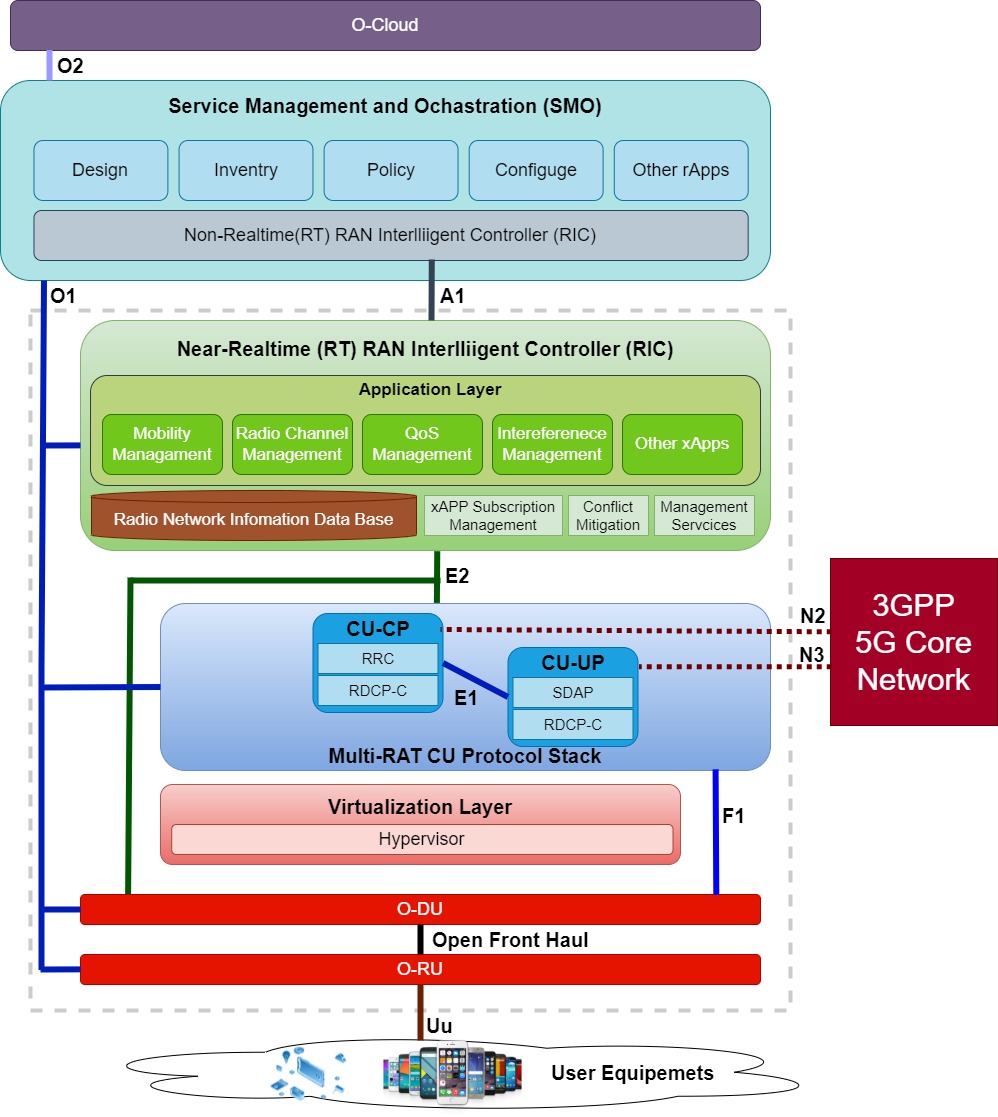}
\caption{The high-level architecture of Open RAN proposed by the O-RAN alliance.} 
\label{Fig:O_RAN}
\end{figure}

O-RAN appliance has proposed a more detailed architecture for Open RAN as represented in Figure~\ref{Fig:O_RAN}. The main elements of the Open RAN architecture include Service Management and Orchestration (SMO), RAN Intelligence Control (RIC), O-Cloud, Open RAN central unit (O-CU), Open RAN distributed unit (O-DU), and Open RAN radio unit (O-RU).

\begin{figure*}
\centering
\includegraphics[width=17.5cm]{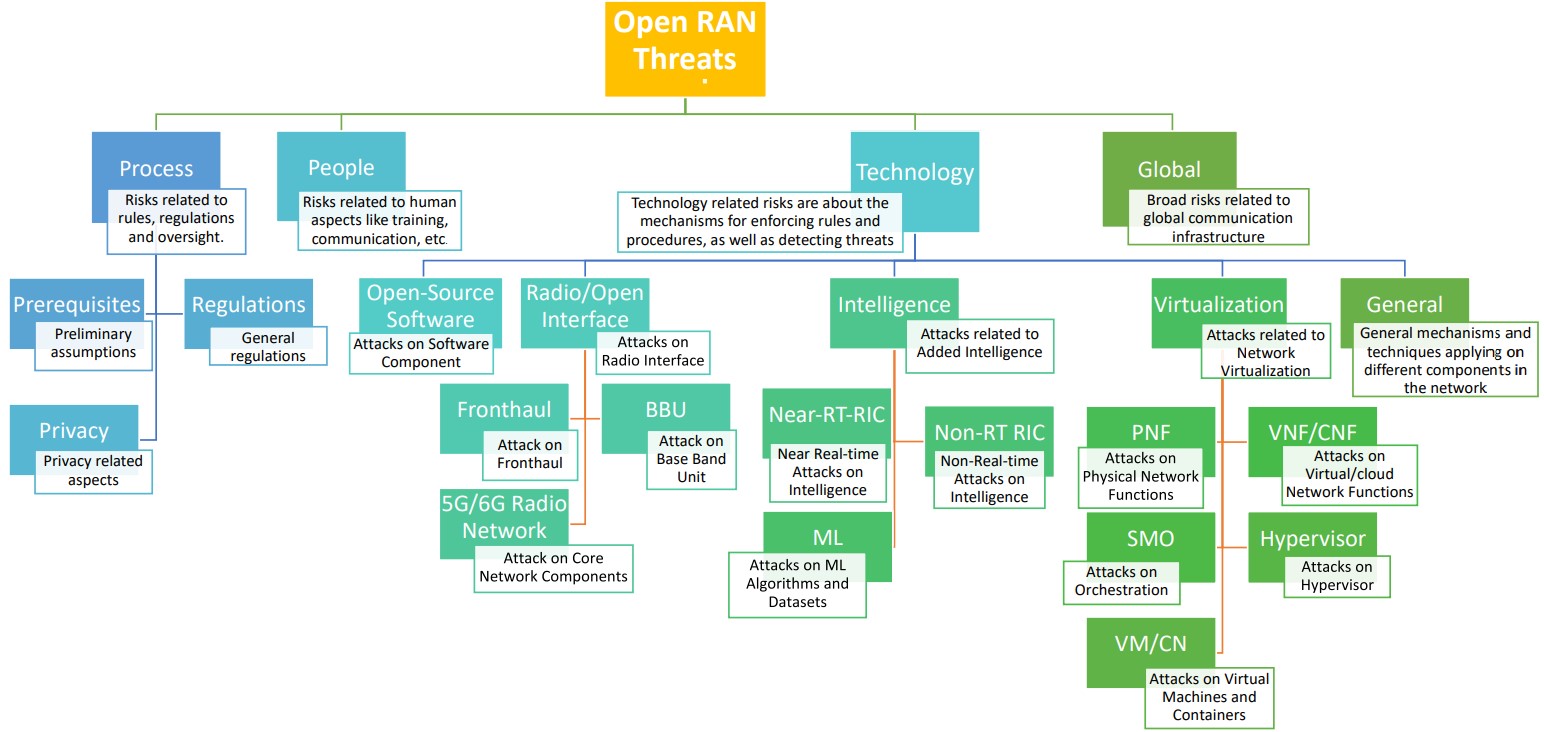}
\caption{\color{black}The Threat Taxonomy of Open RAN Systems\color{black}} 
\label{Fig:risk}
\end{figure*}

\begin{itemize}
    \item \textbf{Service Management and Orchestration (SMO)}: The SMO framework is a core component of the Open RAN architecture, whose main responsibility is to manage the RAN domain, such as the provision of interfaces with network functions, near-real-time RIC for RAN optimization, and O-Cloud computing resource and workload management \cite{wang2021design, wypior2022open}. These SMO services can be performed through four interfaces, including A1, O1, O2, and open fronthaul M-plane.
    
    \item \textbf{RAN Intelligence Control (RIC)}: This logical function enables Open RAN to perform real-time optimization of functions and resources through data collected from the network and end-users. It is the key element in Open RAN, which helps to realize disaggregation strategy, bringing multivendor interoperability, intelligence, agility, and programmability to RANs \cite{balasubramanian2021ric, 7}.   The RIC is divided into components as non-real-time RIC (Non-RT RIC) and near-real-time RIC (Near-RT RIC). The Non-RT RIC is integrated with Open RAN SMO Framework. It handles the control request and RAN resources within the second range. To this task,  Non-RT RIC utilizes specialized applications called rApps. Non-RT RIC can also collect network performance metrics and subscriber data to offer AI-based network optimization and policy guidance recommendations for Near-RT RIC. The Near-RT RIC resides within edge servers or regional cloud as it is responsible for performing network optimization actions within milliseconds range. Near-RT RIC uses the different xApps to support these tasks \cite{orhan2021connection, dryjanski2021toward}. 
    
    \item \textbf{O-Cloud}: This is a physical computing platform. It creates and hosts the various virtual network functions (VNFs) and cloud network functions (CNFs) which are used by near-real-time RIC, O-CU control plane, O-CU user plane, and O-DU\cite{tamim2021downtime}. 
    
    \item \textbf{O-DU}: This logical node has functionalities of the physical and MAC layers. This element terminates the E2 with F1 interfaces.
    
    \item \textbf{O-CU}: This is a logical node in the Open RAN architecture and hosts all the functions of both the control plane and data plane. These two O-CU planes connect with the O-DU logical node via the F1-c interface and F1-u interface, respectively.
    
    \item \textbf{O-RU}: This logical node has a physical layer and radio signal processing capabilities to connect with the SMO framework via the open fronthaul M-plane interface and connects with end-users via radio interfaces.  
\end{itemize}

One of the main goals of Open RAN is ``opening'' the protocols and interfaces between these RAN components, such as radios, hardware, and software. The O-RAN Alliance has defined eleven different interfaces, including A1, O1, E1, F1 open fronthaul M-plane, and O2. More specifically, the open fronthaul M-plane interface is to connect Service Management and Orchestration Framework (SMO) and Open RAN radio unit (O-RU), A1 is to connect non-real-time RAN intelligent controller (RIC) located in the SMO framework and near real-time RIC for RAN optimization, O1 is to support all Open RAN network functions when they are connected with SMO, and O2 is to connect SMO and O-Cloud for providing cloud computing resource and workflow management. According to \cite{garcia2021ran}, there are different deployment scenarios of the O1 interface, such as flat, hierarchical, and hybrid models, by which the SMO framework can provide numerous management services, for example, provisioning management services, trace management services, and performance management services.


\section{Threat Vectors and Security Risks Associated with Open RAN } 
\label{sec:threats}

\color{black}

We start by explaining the taxonomy, used to distinguish the different types of risks. Next, each of the four identified domains are further elaborated.

\color{black}
\subsection{Threat Taxonomy}
 
We categorize the risks in three main domains:  Process, Technology and Global.
First, process risks are related to rules, regulations, oversight. Second, the technology risks correspond with the risks caused by the mechanisms for enforcing rules and procedures, as well as detecting threats. Fourth, global risks are broad risks related to the global communication instruction.	
Figure \ref{Fig:risk}  provides an overview of the risk domains in Open RAN respectively.


\begin{table*}
\caption{Overview of process related Open RAN risks}
\label{prrisks}
\begin{center}
\footnotesize
\begin{tabular}{|p{1.8cm} |p{2cm} |  p{6cm}| p{6cm}| }
\hline
\rowcolor{gray!30}
 Risk category & Threat & Description & Specific to Open-RAN\\
\hline
\hline
Prerequisite	& Requirement of reliable operational environment & \color{black}The operational environment of the Open RAN system must provide reliable timestamps for e.g. the generation of audit records. In addition, the list of minimum prerequisites and assumptions, required to successfully operate the O-RAN system, needs to be defined for the operational environment.  \cite{2,16}  & \color{black} This is applicable to any RAN implementation. However, it is more complex in Open RAN since some aspects (e.g. cloud services) are not under the control of the Open RAN system. \color{black}\\
\cline{2-4}
	& Requirement of secure storage of stored logs, credentials and secrets & Log files, secrets and credentials stored in external systems and related to Open RAN need to be protected, and access controlled should be enable to allow only privileged users. \cite{2,13}  & \color{black}This is applicable to any RAN implementation. However, Open RAN hardware should possess a hardware-based security module like TPM (Trusted Platform Module)  to manage, generate, and securely store cryptographic keys, to offer secure boot, full disk encryption, and remote attestation. \\
	\cline{2-4}
	& Requirement of Trusted certificate authorities (CAs) & Trusted certificate authorities for identity provisioning are applied. \cite{2,13}  & \color{black} This is applicable to any RAN implementation. However, due to involvement of additional stakeholders, the CAs used in Open RAN for authenticating network elements should be properly audited by well established global organizations and SDOs  \\
\hline
General	& Requirement of Secure complete lifecycle process and assessment strategy & Network operators should have an appropriate security process for the complete lifecycle of Open RAN deployment. \cite{2,41}  & \color{black} This is applicable to any RAN implementation. However, it is more complex in Open RAN due the involvement of additional stakeholders.  \\
\cline{2-4}
	& Requirement of Trusted assets/supply chain verification & There is a need to identify, locate, authenticate and verify the origin of the relevant assets in the Open RAN system. \cite{2}  & \color{black} This is applicable to any RAN implementation. However, it is more complex in Open RAN because an Open RAN system is built with components coming from different additional parties.  \\
\cline{2-4}
	& \color{black} Increased complexity and inter-dependency & \color{black} Increased difficulty for identifying issues exists and accountability due to complexity is not evident. \cite{26}  & \color{black} This is specific to Open RAN. Due to the modularity of O-RAN and loss of total ownership. Multiple stakeholders need to collaborate to mitigate the threats  \\
\hline
Privacy	& Violation of privacy policies such as GDPR & Privacy issues arise due to new interfaces, shared environments and new players with different views and objectives on privacy. \cite{2,49} & \color{black} Privacy issues arising from 5G C-RAN are already identified. However, the attack surface increases in case of Open RAN as components can be designed in different regions. \\
\hline

People	& Requirement of Trustworthy and qualified insiders	& There is a need to provide sufficient security resources and sufficient security education and training for the users.	\cite{2,31,44}& The availability of sufficient security educated people is a well-known problem. In the case of Open RAN additional expertise is required such as virtulaized component security \\
\cline{2-4}
&	Requirement of Trusted stakeholders &	All stakeholders involved with Open RAN System should be identified, authenticated and trusted. \cite{2,31}	& This is applicable to any RAN implementation. However, it is more complex in Open RAN due to the increased and diversifed number of stakeholders.  \\

\hline
	
 \end{tabular}
\end{center}
\end{table*}

\subsection{Process}	In the process risks, four categories are distinguished, corresponding to the preliminary assumptions or prerequisites, the general regulations, the privacy  and human related aspects. In fact, all the process risks apply to any RAN implementation, but are in general more complex in Open RAN due to the modularity and the higher amount of stakeholders involved.  Table \ref{prrisks} provides an overview of the key process risks associated with the Open RAN process.

\subsubsection{Prerequisites}
\color{black} To operate a successful Open RAN system, a list of minimum prerequisites and assumptions of the the operational environment needs to be defined. However, The prerequisites  are not under control of the RAN system, but should be carefully checked \cite{ziegler2021make}. To start with, a reliable operational environment must be ensured, providing for instance reliable timestamps to be used in the audit records \cite{palmbach2020artifacts,2}. \color{black}

 \color{black}Next, secure storage of stored logs, credentials and secrets in external systems need to be guaranteed for instance by using hardware based security modules like trusted platform modules (TPMs) \cite{sevincc2007securing,2}. Cryptographic key management, remote attestation, disk image encryption, and secure booting are functions that are typically conducted by a TPM. O-RAN requires such an entity within its midst for managing hardware based security and a root of trust for facilitating signing and verification functions.

In addition, access to this sensitive data should only be allowed by privileged users \cite{2}. The last prerequisite is that the certificate authorities (CAs), which authenticate the network elements, are fully trusted and audited by well established, worldwide recognized organizations \cite{2,dong2016detection}.
In fact, all these prerequisites are essential for any RAN implementation. However, since there are more stakeholders involved in Open RAN, it is clear that these requirements are more challenging to enforce and verify, compared to other RAN implementations.
\color{black}

\subsubsection{General regulations}
The first step in the effective Open RAN launch that needs to be done is the standardization of critical processes like operation, administration and management, covering the complete lifecycle of the Open RAN deployment \cite{kawahara2019ran}. This includes a clear  description of components used for secure establishment of mutual authentication, access control, key management,  trusted communication, storage, boot and self-configuration, update, recoverability and backup, security management of risks in open source components, security assurance, privacy, continuous security development, testing, logging, monitoring and vulnerability handling, robust isolation, physical security, cloud computing and virtualization, and robustness.

Next, it is also important to identify, locate, authenticate and verify the origin of the relevant assets in the system. Furthermore, for each of the different assets, at rest and in transit and location,  the type (data, component, etc.) and the security properties (Confidentiality, Integrity, Availability -  CIA) should be carefully collected. In fact, a complete and efficient supply chain process is required \cite{hassija2020survey}. In particular, this is more complex for Open RAN due to the decoupling of hardware and software and the modularity. For instance, there is a risk of firms from allied states purchasing relabeled products or components from adversarial states. 

Finally, when an issue arises in the network, due to the complexity of the network it is not evident to identify and isolate the issues. Moreover, in case the issue is found, it is possible that the corresponding vendors do not take their responsibility as they can pass the blame to others because of the complexity and interdependence of the whole system.

\subsubsection{Privacy}
The privacy of end users encompasses privacy related to data, identity and  personal information \cite{sorensen20155g}. Privacy sensitive data for end users are mostly leaked via communication services that are gathering all types of personal information, which are often not needed for the functioning of the services. Adversaries can even further extract more personal information about end users, such as User Equipment (UE) priority, location information, trajectory, and preference. The protection of the user data is regulated by the law of the hosting country, where different  jurisdictions can be applicable. There are, at least, three possible locations, the victim, the offender, or the service provider \cite{49}. Therefore, clear guidelines should be developed in order to cope with these new interfaces, shared environments and new players available in Open RAN.

\subsubsection{People}	
First of all, it is necessary to clearly identify and authenticate the stakeholders involved in the different processes like implementation, management, operation and maintenance of the Open RAN system. For each of the stakeholders, their roles and responsibilities should be clearly defined and assessed. Vendors should have well established and transparent security practices built into their engineering processes \cite{2}.

Moreover, adequate training and assessments need to be organized for the different stakeholders, going from administrators, integrators, operators and orchestrators in order to be capable to securely implement and manage the system according to the instructions provided by the Open RAN Alliance and the later to be developed standards \cite{2}. 

Finally, strategies for security testing with published well-known test plans at trusted lab facilities should be defined upfront and integrated in the regular operation \cite{2} Moreover,  adequate training and assessments need to be organized.

\begin{figure*}
\centering
\includegraphics[width=18cm]{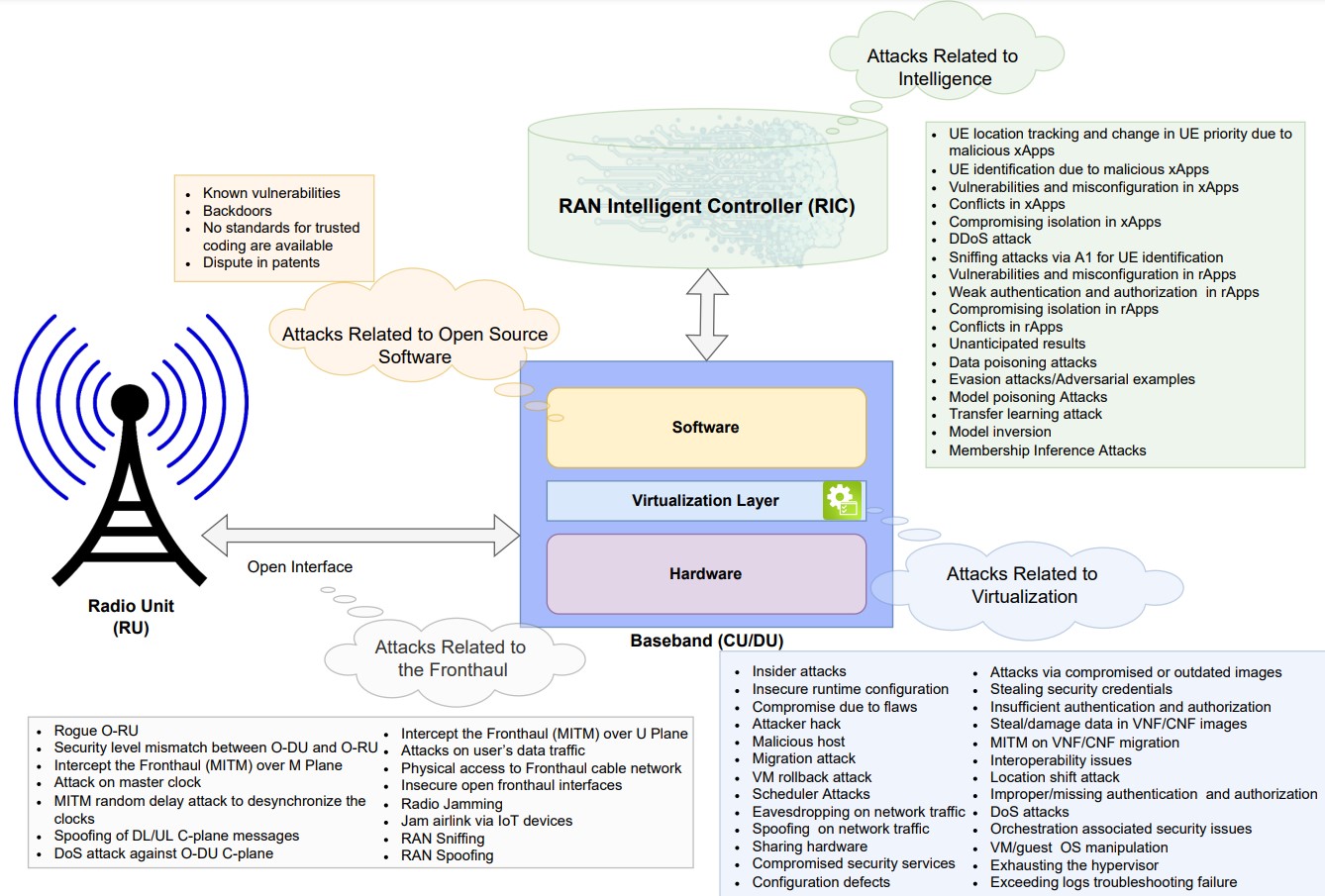}
\caption{\color{black}A Summery of Related Attacks on Open RAN Architecture and It's Components
\color{black}
} 
\label{Fig:techrisk}
\end{figure*}

\subsection{Technology}	
\label{techrisks}
The largest class of risks is related to the different components and mechanisms in the network. Here, distinction is made based on \cite{41}, considering aspects related to open software, radio/open interface, intelligence,  virtualization and general. 
Figure \ref{Fig:techrisk} provides an overview of the main technology related risks in Open RAN.

\begin{table*}
\caption{Overview of Open source software related Open RAN risks}
\label{OSSrisks}
\footnotesize
\begin{center}
\begin{tabular}{|p{2cm} |p{3cm} |  p{6cm}| p{5cm}| }
\hline
\rowcolor{gray!30}
 Impacting Open RAN Component & Threat & Description & Specific to Open-RAN\\
\hline
\hline

All & Known vulnerabilities  &	Attention should be paid to developers using SW components with known vulnerabilities or untrusted libraries and without proper management of interdependencies and patch management. \cite{2,55}	& This is a well-know problem in open source software code. Since Open RAN is expected to be built (solely or partly) based on such open source codes, it is in particular vulnerable for these attacks \\
\cline{2-4}
& Backdoors &	Attention should be paid to a trusted developer intentionally inserting a backdoor into an open source code Open RAN component.\cite{2,15}	& This is a well-know problem in open source software code. Since Open RAN is expected to be built (solely or partly) based on such open source codes, it is in particular vulnerable for these attacks \\ \cline{2-4}
& No standards for trusted coding available	& Explicit legislative standards, guidance, requirements, or conditions to ensure trusted programming should become available.\cite{2,15,55} &	This is a well-know problem in open source software code. Since Open RAN is expected to be built (solely or partly) based on such open source codes, it is in particular vulnerable for these attacks \\ \cline{2-4}
& Dispute in patents  &	The actors involved in Open RAN development implement 5G functions at their discretion and under different copyright regimes. The establishment of a certain type of collaboration is required between those actors as the degree of their collaboration is not at the same level in many cases. \cite{15,28} &	Due to the need of inherent agreements in the O-RAN alliance (With increased number of stakeholders), this is a threat specific for Open RAN. \\

\hline
	
 \end{tabular}
\end{center}
\end{table*}

\subsubsection{Open source software}
The open source related risks are well known problems available in open source software code. An overview is provided in Table~\ref{OSSrisks}. Since, Open RAN is expected to be built (solely or partly) based on such open source codes, it is in particular vulnerable for this type of attacks. 

A trusted developer can intentionally insert a backdoor by injecting a few lines of malicious code into an open source code component to be used within the Open RAN system \cite{li2020backdoor}. It is then highly likely that a software project team picks it up and uses the infected open source code later, while the tools for vetting and testing of the development team do not detect the malicious code \cite{2}. As a consequence, a vulnerability into the software code is included and can go undetected for a long period. The resulting effect on the Open RAN system can be diverse. It can either be simply annoying, but at the same time it can significantly decrease the system performance via for instance Denial of Service (DoS) attacks, or it can even lead to serious loss of sensitive data.

Open source vulnerabilities are normally published on the National Vulnerability Database (NVD) \cite{booth2013national}. This database is primarily intended for developers to disclose vulnerabilities. However, this source is also used by hackers to exploit those vulnerabilities enabling backdoors to attacks on e.g. the hypervisor, Operating System (OS), Virtual Machine (VM) or container. Moreover, vulnerabilities frequently propagate as developers often re-use free open source code. As a consequence, downloading open source libraries and their dependencies, as well as downloading open source code from untrusted repositories contain significant risks \cite{2}. Open RAN vendors and operators should thus store at each moment up-to-date inventories containing the dependencies in their open source software used in the applications. In addition, this should be  complemented by a process, which receives and manages all the notifications coming from the open source community that are related to newly discovered vulnerabilities, including newly developed patches to overcome them. This should enable a better supply chain traceability.

Existing legislation demonstrates implied security preferences but provides no explicit legislative standards, guidance, requirements, or conditions. These preferences should be explicit but transparent, reviewable, and auditable to ensure secure coding. Due to the fact a material amount of Open RAN code is being written by firms in different countries, security audits should be mandatory making code available to security researchers  \cite{15}. 

Finally, the last open source software risk is more linked to political and financial interests, instead of security interests.  Both the 3GPP and Open RAN alliance operate a  Fair, Reasonable and Non Discriminatory (FRANS) policy when it comes to patents that are held by contributors to those respective organizations. Patents are held on aspects of the 3GPP and Open RAN Alliance specifications, but the holders of those patents agree that it is mutually beneficial for everyone if the patents are licensed with a FRANS approach. The concern in this area is politically oriented. There might be a possibility that the patents held by competing manufacturers and service providers may be withdrawn from the FRANS licensing arrangement  if trade relations between different countries dramatically deteriorate  \cite{15,mariniello2011fair}.       

\subsubsection{Radio/Open Interface}
The different radio/open interface components include the Fronthaul, the central Unit/distributed unit (CU/DU) and the 5G radio network. \color{black}Table \ref{RadioSrisks} summarizes the threats related to this radio/open interface.
\color{black}
\begin{table*}
\caption{Overview of Radio/Open Interface related Open RAN risks}
\label{RadioSrisks}
\footnotesize
\begin{center}
\begin{tabular}{|p{1.5cm}|p{2cm}|p{7.5cm}|p{5cm}|}
\hline
\rowcolor{gray!30}
 Impacting Open RAN Component & Threat & Description & Specific to Open-RAN\\
\hline
\hline

Fronthaul &	Rogue O-RU	&The idea is to fool O-DU or UE into associating it to a rogue O-RU over the legitimate O-RUs. \cite{2}	& It is possible to set up a rogue RU  in other RAN systems as well. However, it will be more easy to develop a rogue O-RU in O-RAN due to the open nature. No implicit security due to lack of knowhow. 
\\
\cline{2-4}
	&Security level mismatch between O-DU and O-RU	&An attacker penetrates O-DU and beyond through O-RU or the Fronthaul interface due to heterogeneous security levels in the split architecture. \cite{2,5}	&Yes, This is happen only in O-RAN as different vendor equipment is possible for O-RU and O-DU.
\\
\cline{2-4}
&	Intercept the Fronthaul (MITM) over M Plane	&The high bit rate Fronthaul interface imposes strict performance requirements which force to limit the use of some security features. \cite{2}	&Yes, This interface is specific in O-RAN.
\\
\cline{2-4}
	&Attack on master clock	&An attacker can attack a master clock by sending an enormous amount of time protocol packets. It can also impersonate a legitimate clock, a slave, or an intermediate clock, by sending malicious messages to the master, thus degrading the victim's performance. \cite{2,dik2021transport}	&\color{black} No, however, there is an increased range of attacks in Open RAN due to the use of various xApps and rApps. Moreover, near-RT operation is expected in Open RAN for some of the functions.
\\
\cline{2-4}
	&MITM random delay attack to desynchronize the clocks	&An attacker acting as MITM can introduce random packet delay on  Precision Timing Protocol (PTP) sync messages and/or PTP delay-req/resp messages, which causes inaccurate PTP offset calculation, thus the clocks may not be synchronized properly. \cite{2,dik2021transport}	&No, however, there is an increased range of attack  in O-RAN.
\\
\cline{2-4}
	&Spoofing of DL/UL C-plane messages	&\color{black}An adversary  injects DL/UL C-plane messages that falsely claim to be from the associated O-DU which would impact the O-RU to process the corresponding U-Plane packets \cite{2,9}. This will  lead to temporarily limited cell performance (or even DoS) on cells served by the O-RU and in addition, a consequential threat to all O-RUs parented to that O-DU might exist. &\color{black}	No. However, there is an increased range of attacks in Open RAN. Moreover, this attack can be easier to perform  in shared virtualized environment. 
\\
\cline{2-4}
	&DoS attack against O-DU C-plane	& \color{black} DoS attacks against the O-DU C-plane are launched. \cite{2}. Due to the cleartext nature of Enhanced Common Public Radio Interface (eCPRI) messages used for the Open Fronthaul C-Plane, an attacker can launch a volumetric DoS attack with bad or unauthenticated eCPRI Real-time control data messages (adopted for C-Plane communication) against the O-DU C-Plane, causing O-DU performance degradation and potentially its overall service interruption, which could further cascade to all its serving O-RUs.	& \color{black} No, however, there is an increased range of attacks in Open RAN. The openness in the Open RAN system will be a cause to lose explicit security due to a lack of know-how. 
\\
\cline{2-4}
	&Intercept the Fronthaul (MITM) over U Plane&	An attacker attempts to intercept the Fronthaul (MITM) over the User Plane due to the limited use of some security features at the Fronthaul interfaces. \cite{2}	&No, This is a problem also in 3GPP RAN.
\\
\cline{2-4}
&	Attacks on user's data traffic	&The integrity protection is enabled on the Control Plane messages, which still makes the data traffic of the user vulnerable because the Control Plane and User Plane are segregated. \cite{2,5} & \color{black}	No, This is a problem also in 3GPP RAN. However, Open RAN offers the computing resources (i.e. not available in other RANs) to implement  3GPP specified UP integrity protection algorithms without impacting on the user experience.
\\
\cline{2-4}
	&Physical access to Fronthaul cable network 	&An intruder into the exchange over the Fronthaul cable network attempts to gain electronic access to cause damage or access sensitive data. \cite{2,bitsikas2021don}	&The same type of attack can be applied in other RAN. However, the attack range and possibilities are increased in O-RAN.
\\
\cline{2-4}
	&Insecure open Fronthaul interfaces	&An attacker penetrates and compromises the O-RAN system through the open O-RAN's Fronthaul due to the lack of industry level security best practices.	\cite{2,9,37} &Yes, since this involves the new interfaces introduced specific in O-RAN. Although following \cite{37}, it is also inherently required for a secure 5G implementation. \\ 
	\hline
CU/DU	&  Attacks via Shared Baseband Units	& \color{black} An attacker exploits the lack of isolation on Shared Baseband Units and the edge platforms to perform attacks. \cite{25,ranaweera2021survey} \color{black} & The same type of attack can be applied in V-RAN. However, the attack range and possibilities are increased in O-RAN. \\ 
	\hline
5G Radio Network	&Radio Jamming	&An attacker could disrupt the communication by deliberate jamming, blocking or creating interference with the authorized wireless network. \cite{2}&	No, this is a general attack that can be applied to any RAN.
\\
\cline{2-4}
	&Jam airlink via IoT devices&	An attacker attempts to jam the airlink signal through IoT devices. \cite{2}	&No, this is applicable to any RAN.
\\
\cline{2-4}
	&RAN Sniffing 	&An attacker could decode the essential network configuration details by sniffing the RAN. \cite{2}&	No, this is a general attack that can be appied to any RAN.
\\
\cline{2-4}
	&RAN Spoofing &	An attacker is spoofing the RAN signals by transmitting a fake signal meant to pretend as an actual signal.	\cite{2}&No, this is a general attack that can be applied to any RAN. \\ 
\hline
	
 \end{tabular}
\end{center}
\end{table*}

\begin{itemize}
\item \textbf{Fronthaul}.
The Fronthaul of Open RAN, consisting of O1, O2, A1, and E2 are the new components, all available with open interfaces allowing the software programmability of RAN. These components and interfaces may not be secured to industry best practices, for instance containing no proper authentication and authorization processes, ciphering and integrity checks, protection against replay attacks, prevention of key reuse, validation of inputs, response to error conditions, etc \cite{2}. This follows often from the  strict performance requirements (bandwidth, latency, fronthaul transport link length, etc.) that limit the use of some security features, enforced by the high bit rate fronthaul interface to increase the processing delay. As a consequence, different MITM, DoS, data tampering or even information disclosure attacks become possible.

\color{black}
\begin{itemize} 
\item The first category of risks is due to attacks from the internet exploiting weak authentication and access control to penetrate the network boundary. There are several possibilities for this.

First, it would allow the presence of a rogue Open RAN Radio Unit (O-RU) in order to fool the O-DU or UE into associating with it instead of the legitimate O-RUs \cite{2}.  This opens the door to subscriber's identity interception/disclosure and unauthorized user tracking attacks of user movements and activities by catching the SUPI/5G-GUTI of the subscriber's User Equipment (UE) and location of a device).

Second, an adversary can inject DL/UL C-plane messages that falsely claim to be from the associated O-DU, which would impact the O-RU to process the corresponding U-Plane packets \cite{9}. Also spoofing of DL/UL C-plane messages, leading to temporarily limited cell performance (or even DoS) on cells served by the O-RU and in addition, a consequential threat to all O-RUs parented to that O-DU might exist.  

Third, if in addition no trusted stakeholders are guaranteed, an attacker can attack a master clock by sending an excessive amount of time protocol packets or impersonate a legitimate clock, a slave, or an intermediate clock, by sending malicious messages to the master, thus degrading the victim's performance \cite{dik2021transport}. The attacker may be residing either within the attacked network (insider) or on an external network connected to the attacked network. This attack results in a situation where the clock service is interrupted completely or the timing protocol is operational but slaves are being provided inaccurate timing information due to the degraded performance of the master clock. This degradation in the accuracy of time may cause DoS to applications on all the RUs that rely on accurate time, potentially bringing down the cell. A cell outage caused by misaligned time may further impact performance in connected neighboring cells.

Finally, when having two different vendors, the O-RU and the O-DU need to be managed as different entities and may have heterogeneous security levels \cite{5}. Instead, the O-DU will have to bridge the management traffic between the management system and the O-RU. Hence the possibility to reach the northbound systems beyond the O-DU through the Open Fronthaul interface becomes a possible attack vector in this split architecture.

\item The second category of risks on the fronthaul is due to the ability of the attacker to compromise Open RAN data integrity, confidentiality and traceability in case the components are not secured to the industry best practices. This follows often from the  strict performance requirements (bandwidth, latency, fronthaul transport link length, etc.) that limit the use of some security features, enforced by the high bit rate fronthaul interface to increase the processing delay. As a consequence, different MITM attacks become possible.

A MITM attacker over the fronthaul interface is able to intercept the data over the U-Plane and introduce random packet delay on the Precision Timing Protocol (PTP) sync messages and/or PTP delay-request/response messages, which causes inaccurate PTP offset calculation, resulting in clocks which may not be synchronized properly \cite{dik2021transport}. Also, denial of service (DoS) attacks become possible. Moreover, after breaking the PDCP security, also access to content can be obtained.

A Man-in-the-Middle (MITM) attack over the fronthaul interface or O1 is able to intercept the M plane, and thus also to do  passive wiretapping and DoS, but needs to break M Plane Security prior to gain OAM access \cite{2}.

\item The third category of risks on the fronthaul is if an attacker compromises the Open RAN monitoring mechanisms and integrity and availability of the log files \cite{sasaki2020security}.

\item The fourth category of risks on the fronthaul is caused by a compromise of the integrity and availability of the Open RAN components in general. Insufficient assurance of Open RAN software package integrity could affect CIA of data, services, hardware and policies during installation or upgrade phases for Open RAN components \cite{2}. An attacker could, in such a case, cause denial-of-service, data tampering, information disclosure, spoofing identity, etc.

\item Finally, if an attacker is able to get physical access to the fronthaul components, it can result in a devastating impact on the confidentiality and integrity of the data \cite{bitsikas2021don}. Note that this is typically linked to the first type of process related threats dealing with trusted stakeholders.

\end{itemize}
\color{black}


\item 	\textbf{CU/DU}. The shared units pool in the Open RAN cloud native deployment may suffer from insufficient isolation and impose the risk of breaking user privacy and accessing sensitive data \cite{25}. \color{black} The openness and exposure of the CU and DU entities in comparison to C-RAN are inviting intruders for gaining access of those entities through cyber hacking attempts. As the fronthaul of the O-RAN is expected to be deployed via enhanced Common Public Radio Interface (eCPRI), converged packet based network that contrives it is inviting cyber threats unlike the traditional frounthauls \cite{kazemifard2021minimum}. Although uncommon, intrusions can be perpetrated via the F interface in the Mid-haul that connect CU to its corresponding DUs. Such interventions are possible through the threat vectors such as service migration, offloading, or handover mechanisms that exist with edge computing base stations that are presumed to host CUs \cite{ranaweera2021survey}. A compromised CU is capable of impregnating both fronthaul and the backhaul directions leveraging the open interfaces of the O-RAN. \color{black}

\item 	\textbf{5G Radio Network}.
These attacks are classical attacks, which can be applied to any RAN system and include radio jamming, jamming via IoT devices, RAN sniffing and spoofing \cite{2}. 

Radio jamming can be impacting on the reference signals, the synchronization signal, the Physical Broadcast Channel (PBCH), the Physical Downlink Control Channel (PDCD), the Physical Uplink Control Channel, or the Physical Random-Access Channel \cite{chi2020countering}. This would enable an attacker to disrupt the communication by deliberate jamming, blocking or creating interference with the authorized wireless network. \color{black} Additionally to blocking the communication flow, jamming the synchronization channels or the signaling flow is another method to disrupt the 5G services \cite{varga20225g}. A capable adversary can target different entities of the 5G communication network simultaneously to impact an interference significant enough to subdue the communication. Thus, a jamming detection mechanism is mandatory to filter out the jamming frequencies in this era of 5G and beyond \cite{wang2022anonymous}. \color{black}

In addition, due to the millions of IoT devices in the network, jamming of the airlink signals through the IoT devices, can easily overload the Open RAN resources by means of Distributed DoS (DDoS) attempts carried via a botnet army of millions to billions of infected devices, on which a malware instructs to reboot all devices in a specific or targeted 5G coverage area at the same time \cite{wood2004taxonomy}. \color{black} Most IoT based services are Location Based Services (LBSs) and expect locational awareness with utmost availability. The attackers capable of jamming the GPS receiver will succeed in subduing the service to an inaccurate state \cite{varga20225g}. Since the O-RAN interfaces should be open to a common standardization to avoid vendor-specific nature, adversaries have the ability to assimilate the firmware and software specifications and induce a race-like condition by exploiting its vulnerabilities. \color{black} 

\color{black} RAN sniffing allows the attacker to decode essential network configuration details, assisting attackers to optimize and craft their attacks \cite{alina2021understanding}. Vulnerabilities of the PBCH channel are allowing the attackers to sniff the 5G RAN network stats \cite{he2018lte}. The open-source and low-cost natures of the software radio are inviting the attackers to exploit the existing vulnerabilities in software, protocol, and firmware layers. With RAN spoofing a fake signal pretending to be an actual signal is conveyed by targeting an RF receiver within the RAN \cite{alina2021understanding}. Similar to sniffing, vulnerabilities of the PBCH channel and the software radio devices can be the main causes targeted through spoofing attempts, that embrace the masquerading signal as a legitimate one \cite{lichtman20185g}. \color{black}

\end{itemize}

\subsubsection{Intelligence}	

\begin{table*}[htbp]
\caption{Overview of Intelligence related Open RAN risks}
\label{AIrisks}
\scriptsize
\begin{center}
\begin{tabular}{|p{1.5cm} |p{2.5cm} |  p{9cm}| p{2.5cm}| }
\hline
\rowcolor{gray!30}
 Impacting Open RAN Component & Threat & Description & Specific to Open RAN\\
\hline
\hline

Near-RT-RIC	&	UE location tracking and change in UE priority due to malicious xApps	&	xApps have the capability to manipulate behavior of a certain cell, a group of UEs, and a specific UE to track a certain subscriber or change priority level of an UE. \cite{2,5}	&	Yes since xApps and E2, A1 interfaces are only defined in O-RAN.	\\ \cline{2-4}
	&	UE identification due to malicious xApps	& \color{black}	Malicious xApps can exploit UE identification and track UE location. For example, a xApp can potentially be used as a ``sniffer'' for UE identification. \cite{2,5,9}	&	Yes since xApps and E2, A1 interfaces are only defined in O-RAN.	\\ \cline{2-4}
	&	Vulnerabilities and misconfiguration in xApps	& \color{black} Vulnerabilities can potentially exist in any xApp, if it obtained from an untrusted or unmaintained source. An attacker exploits  vulnerabilities and misconfiguration of such xAPPs to disrupt the offered network service and potentially take over another xApp or the whole near-RT RIC. \cite{2,5,abdalla2021toward}	&	Yes, as these components are only defined in O-RAN.	\\ \cline{2-4}
	&	Conflicts in xApps	&\color{black}	Conflicting xApps unintentionally or maliciously impact O-RAN system functions such as mobility management, admission controls, bandwidth management and load balancing in the purpose of performance degradation. Moreover, a threat actor can utilize a malicious xApp that intentionally triggers RRM (Radio Resource Management) decisions conflicting with the O-gNB internal decisions to create DoS.
  \cite{2,5}	&	Yes, as these components are only defined in O-RAN.	\\ \cline{2-4}
	&	Compromising isolation in xApps	&\color{black}	An attacker compromises xApp isolation to break out of xApp confinement. Such a way, attacker can perform side channel attack to deduce information from co-hosted xApps in  a shared resource pool. \cite{2} &	Yes, as these components are only defined in O-RAN.	\\ \hline
Non-RT RIC	&	DDoS attack 	&	An attacker penetrates the Non Real-Time RAN Intelligent Controller (Non-RT RIC)  to cause a DoS or degrade the performance. \cite{2}	&	Yes, as these components are only defined in O-RAN.	\\ \cline{2-4}
	&	Sniffing attacks via A1 for UE identification	& \color{black}	An attacker performs UE sniffing in the Non-RT RIC via A1 interface or via R1 interface via rApps in order to identify UE.  For example, a rApp can potentially be used as a ``sniffer'' for UE identification.  \cite{2}	&	Yes, as these components are only defined in O-RAN.	\\ \cline{2-4}
	&	Vulnerabilities and misconfiguration in rApps	&	\color{black} Vulnerabilities can potentially exist in any rApp, if it obtained from an untrusted or unmaintained source. An attacker exploits  vulnerabilities and misconfiguration of such rAPPs to disrupt the offered network service and potentially take over another rApp or the whole non-RT RIC. \cite{2,5,abdalla2021toward}	&	Yes, as these components are only defined in O-RAN.	\\ \cline{2-4}
	&	Weak authentication and authorization in in rApps	& \color{black}	If web front-end or REST API  interfaces contain software vulnerabilities or implement authentication and authorization insufficiently, an attacker could bypasses authentication and authorization and able to gain access to the rApp and pose as a tenant. Such a way an attacker gains the ability  manipulate configurations, access logs and implement back doors \cite{2,9} &	Yes, as these components are only defined in O-RAN.	\\ \cline{2-4}
	&	Compromising isolation in rApps	&	\color{black}	An attacker compromises rApp isolation to break out of rApp confinement. Such a way, attacker can perform side channel attack to deduce information from co-hosted rApps in  a shared resource pool\cite{2}	&	Yes, as these components are only defined in O-RAN.	\\ \cline{2-4}
	&	Conflicts in rApps	&\color{black}	Conflicting rApps (i.e. direct, indirect and implicit conflicts) unintentionally or maliciously impact non-realtime Open RAN system functions such as Carrier license scheduling, energy savings and subscription handling to degrade performance or trigger a DoS.  \cite{2}	&	Yes, as these components are only defined in O-RAN.	\\ \hline
ML	&	Unanticipated results	& \color{black}	If unexplainable AI is used, the results cannot be predicted and might have a fast impact \cite{yampolskiy2020unexplainability}. Therefore, use of unexplainable  AI/ML models in the Open-RAN can potentially lead to unanticipated consequences, which might have an impact on the security and privacy \cite{9}. Such AI/ML models could unintentionally violate the security and privacy policies and offer bias results. &	The same type of attack can be applied in V-RAN. However, the attack range and possibilities are larger in O-RAN.	\\ \cline{2-4}
	&	Data poisoning attacks 	& \color{black}	An attacker with access to the training set is able to poison the ML training data (Data poisoning attacks) and thus break the reliability of the training. This impacts the xApps/rApps managed Open RAN system functions  such as mobility management, admission controls, bandwidth management, load balancing and results a performance degradation \cite{2,sun2021data}.	&	This attack only applies to O-RAN, where ML is explicitly included.	\\ \cline{2-4}
	&	Evasion attacks/Adversarial examples	&\color{black}	An attacker uses an adversarial example (intentionally designed date) as an input to the ML models to make a mistake\cite{61,64,iturria2022multi}. This impacts the xApps/rApps managed Open RAN system functions   results a performance degradation. &	This attack only applies to O-RAN, where ML is explicitly included.	\\ \cline{2-4}
	&	Model poisoning Attacks 	& \color{black}	An attacker with access to the model can alter the ML model resulting in system manipulation and compromise of ML data confidentiality and privacy \cite{2,iturria2022multi,shi2021adversarial}. This impacts the xApps/rApps managed Open RAN system functions   results a performance degradation.	&	This attack only applies to O-RAN, where ML is explicitly included.	\\ \cline{2-4}
	&	Transfer learning attack 	& \color{black}	A transfer learning attack becomes possible \cite{2}. This impacts the xApps/rApps managed Open RAN system functions   results a performance degradation.	&	This attack only applies to O-RAN, where ML is explicitly included.	\\ \cline{2-4}
	&	Model inversion	& \color{black}	An attacker can have the aim to reconstruct training data from model parameters	\cite{61,62,63}. This impacts the xApps/rApps managed Open RAN system functions   results a performance degradation.&	This attack only applies to O-RAN, where ML is explicitly included.	\\ \cline{2-4}
	&	Membership Inference Attacks 	& \color{black}	An attacker tries to identify the data samples used for the model training	\cite{61,65}. This impacts the xApps/rApps managed Open RAN system functions   results a performance degradation.&	This attack only applies to O-RAN, where ML is explicitly included.	\\

\hline
	
 \end{tabular}
\end{center}
\end{table*}

The different components and mechanisms that contribute to the intelligence in the Open RAN network are the Near Realtime Radio Access Network Intelligence Controller (Near-RT-RIC), Non Realtime RIC (Non-RT RIC), and machine learning (ML) algorithms. These risks are mostly specific to Open RAN as they operate on new components and new algorithms, which are currently not available. \color{black}The threats related to intelligence are summarized in Table \ref{AIrisks}.\color{black}


\begin{itemize}
\item \textbf{Near-RT-RIC related Attacks:}
xApps have the capability to manipulate behavior of a certain cell, a group of UEs, and a specific UE. The related attacks are due to either malicious xApps, xApps with vulnerabilities, misconfigured xApps, compromised xApps or conflicting xApps \cite{2}.

\color{black} As xApps are launched to perform intelligent functions for CU and DU entities in regards to radio resource management, a compromised xApp could attempt to take the control of a cell, a RU device, or a group of UE devices; and would be capable of tracking a certain consumer within its RIC domain. In addition, the same malicious xApp could gather priority information on the served UE devices through the A1 interface, where distinguishing and identifying serving UEs are possible. Such acts violate the location privacy of the important UEs and even the prioritization on the currently serving services can be manipulated. \color{black} This will leads to compromise RAN performance as well as the privacy violation.

Malicious xApps can potentially be used as a sniffer for UE identification. \color{black} In such a circumstance, RAN performance could be impacted negatively while the privacy of the subscribers may be violated. \color{black} This follows from the fact that the A1 interface is able to point out a certain UE in the network (through its UE identifier), which creates correlations among the randomized and anonymized UE identities between the RAN nodes. As a consequence, UE location tracking and change in UE priority become possible. In particular, identification and tracking of a certain subscriber, for instance, a Very Important Person (VIP) becomes a real threat. \color{black} The exposure of the UE identifier is most probable through E2 signaling channels in comparison to its counterpart A1, due to the Near-RT conditions of the E2. Further, such malicious xApps could change the Service Level Agreement (SLA) specifications of the assigned services similar to changing of the priority levels. Such acts could conflict with the Near-RT-RICs decision process as the program execution times might extend beyond the specified boundaries of a presumed Near-RT event or SLAs \cite{abdalla2021toward}. \color{black}

Vulnerabilities can potentially exist in any xApp, since it can come from either an untrusted or unmaintained source. Such vulnerabilities can then be exploited to take over another xApp or the whole near-RT RIC and often have the purpose to degrade the performance (e.g. a DoS). It may also be possible to alter data transmitted over A1 or E2 interfaces, or to extract sensitive information. Also, the xApp isolation can be exploited in order to break out of the xApp confinement and to deduce information from co-hosted xApps. In addition, unauthorized access provides new opportunities to exploit vulnerabilities in other xApps or Open RAN components to intercept and spoof network traffic, and to degrade services (through DoS). \color{black} The open source nature of the xApps are advertising the vulnerabilities to the adversaries while misconfigurations and incompatibilities are inevitable with the open nature of the O-RAN. \color{black}

Finally, since there is no clear functional split between the Near-RT RIC and the Open RAN Next Generation Node B (O-gNB), possible conflicts, including conflicts in xApps, between the decisions taken by the Near-RT RIC and the O-gNB regarding the radio resource management can appear, both unintentionally or maliciously. This can have an impact on the Open RAN system functions such as mobility management, admission controls, bandwidth management and load balancing, potentially resulting in performance degradation. \color{black} Moreover, isolation of xApps is critical for the independent operation of O-RAN services and for accurate Near-RT-RIC decision making. Such an isolation or confinement can be penetrated through underlying system vulnerabilities, deducing access information via shared resource applications, or masqueraded authentication attempts. A compromised isolation could subdue the xApp operations to the attacker. \color{black}

\item 	\textbf{Non-RT RIC related Attacks:}
\color{black} rApps impact non-RT RIC functions such as AI/ML model training, A1 policy management, enrichment information management, network configuration optimization for the purpose of performance degradation, DoS, and enrichment data sniffing (UE location, trajectory, navigation information, and GPS data). rApps bearing many resemblances to xApps in their operational context, have the ability to manipulate the behavior of a certain cell, a group of UEs, and a specific UE. The related attacks are similar to xApps, due to either malicious rApps, rApps with vulnerabilities, misconfigured rApps, compromised rApps or conflicting rAPPs. Besides these similar ones, there are two more risks identified related to Non-RT RIC \cite{2}.

Untrusted or unmaintained sources can cause vulnerabilities in any rApp. Exploitation of these vulnerabilities mostly leads to disruption of the offered network service and potentially taking over another rApp or the whole non-RT RIC. As a consequence, the attacker may gain the ability to alter data transmitted over A1 interface, or extract sensitive information. Also rApp isolation can be exploited to break out of rApp confinement and to deduce information from co-hosted rApps. Unauthorized access provides new opportunities to exploit vulnerabilities in other rApps or Open RAN components to intercept and spoof network traffic, to degrade services through DoS attempts; An attacker might also penetrate the non-RT RIC through A1/O1 interfaces or from external sources through SMO and attempts to trigger a DoS or degrade the performance of non-RT RIC \cite{abdalla2021toward}.

In addition, rApps in the Non-RT RIC can cause conflicting decisions as they can be launched by different vendors targeting different purposes: Carrier license scheduling, or energy savings. Such conflicts could take the form of a direct, indirect, or implicit nature depending on the rApp parameters in question, and the effect that particular conflict is inducing. As in direct ones deals with the conflict of same parameter change requested by different rApps, indirect ones where the different parameter changes by different rApps would cause an opposite effect, and implicit ones that different parameter changes would lead to changing the network state. The effects can lead to an overall network performance degradation, or instabilities within the network entities. These conflicts are difficult to mitigate since dependencies are impossible to observe.

There is an additional vulnerability that can appear in the case the rApp management is exposed to a web front-end or REST API, whose software interfaces contain vulnerabilities or do not implement authentication and authorization in a proper way. 
This would allow  an attacker to gain access to the rApp and pose as a tenant or to manipulate configurations, access logs, or to implement back doors.

\color{black}

\item 	\textbf{ML related Attacks:}
\color{black} ML and AI play a vital role in the formation of the O-RAN concept. Thus, vulnerabilities or flows in existing ML models or algorithms can be envisaged as probable threats to the O-RAN system as they are deployed in the intelligence portion of the architecture. One of the most common threats is the data poisoning attacks, where the adversary is altering the data sets that are intended for training, testing, or validating \cite{sun2021data,2}. The access to perform such modifications, however, can be gained via the penetration through fronthaul, midhaul, xApps, or rApps. Poisoning attempts could impact any stage of the ML process as in feature selection, prediction, decision making, model classification, or anomalous detection. The O-RANs openness and the Near-RT operations require the ML models to be formed online with continual updating during the operation. Though it would not impact in the long run, feeding bogus data to the online ML model is capable of impacting the RIC decision making negatively, especially in terms of radio resource allocation. Similarly, evasion/ adversarial attacks or model poisoning attacks represent two variants of the poisoning attack. In the evasion attempt, data is carefully tampered according to a perturbed design that would not detect as anomalous. In the model poisoning attempt, the entire model or the control parameters of the model are altered to impact the learning phases of the process \cite{shi2021adversarial}. 

Pre-trained and widely available ML models can be utilized by the attackers for gaining access or evading the system's anomalous detectors for launching transfer learning attacks. Model inversion and membership inference attacks are ensuing privacy leakages \cite{61}. In model inversion attempts, adversary is reconstructing the training data set from the model parameters \cite{62,63,benzaid2020ai}. This is plausible as there are plenty of online repositories with training data that would aid the attacker in cross-validating the determined data. Membership inference threat would determine whether a particular data set was used in the training process of the ML model or not \cite{65}.

Finally, due to the complexity of the models in AI/ML, the results are not yet explainable in most of the cases \cite{yampolskiy2020unexplainability}. Therefore, its use in the RAN can potentially lead to unanticipated consequences, which might have an impact on the security or performance \cite{9}.


\color{black} The impact of data poisoning attempts would clearly target the allocation and management of radio resources within the O-RAN fronthaul, and could result in jeopardizing the accuracy of mobility management, load balancing, and QoS management functions that are administered under the Near-RT-RIC. In the long run, the entire RT intelligent framework could become compromised. Model poisoning, evasion and transfer learning attempts induce the same impact on the RT systems. For the Non-RT system, however, as the time is not a critical parameter, the impact would be less costly, as the decisions are made from the data gathered from an extended period in comparison to the RT instances.

\color{black}
\end{itemize}

\subsubsection{Virtualization}

\begin{table*}
\caption{Overview of Virtualization related Open RAN risks}
\label{VNrisks}
\footnotesize
\begin{center}
\begin{tabular}{|p{1.4cm} |p{3cm} |  p{7cm}| p{4cm}| }
\hline
\rowcolor{gray!30}
 Impacting Component & Threat & Description & Specific to Open-RAN\\
\hline
\hline

PNF-VNF/CNF	&	Lack of security policies to protect mixed PNF-VNF/CNF	&	Compromises a PNF to launch reverse attacks and other attacks against VNFs/CNFs due to the lack of security policies to protect mixed PNF-VNF/CNF\cite{2,benzaid2020ai}.	& Can be applied in V-RAN. However, the attack possibilities are increased in Open RAN.		\\ \hline
VNF/CNF	&	Attacks via compromised or outdated images 	&	Compromises VNF/CNF images or used outdated images\cite{2}	&	The same type of attack can be applied in V-RAN. 	\\ \cline{2-3}
	&	Configuration defects 	&	Utilizes the configuration defect of VNF/CNF to attack\cite{2}.	& However, the attack range and possibilities are increased in Open RAN.		\\ \cline{2-3}
	&	Stealing security credentials 	&	Steals embedded security credentials from VNF/CNF images\cite{2}.	&		\\ \cline{2-3}
	&	Insufficient authentication and authorization 	&	Insufficient authentication and authorization can lead to IP loss and expose significant technical details about an Open RAN VNF/CNF image to an attacker.	\cite{2}&		\\ \cline{2-3}
	&	Stealing or damage of embedded information from VNF/CNF images 	&	Steal or damage sensitive information from/in VNF/CNF images\cite{2}.	&		\\ \cline{2-3}
	&	MITM on VNF/CNF migration	&	Performs MITM to intercept network traffic and jeopardize the VNF/CNF image migration. \cite{2,ranaweera2021survey}	&		\\ \cline{2-3}
	&	Interoperability issues	&	Exploits the security level mismatches of different VNF/CNFs\cite{50,49,ranaweera2021survey,lal2017nfv}.	&		\\ \cline{2-3}
	&	Location shift attack	&	A compromised VNF/CNF changes the run-time environments to perform an attack\cite{50,lal2017nfv,ranaweera2021survey}	&		\\ \hline
SMO	&	Improper/missing authentication 	&	Can exploit the improper/missing authentication on Service Management and Orchestrator (SMO) functions to illegally access the SMO and its functions\cite{2}. &	The same type of attack can be applied against the Management and Orchestration (EMM) in C-RAN. 	\\ \cline{2-3}
	&	Improper/missing authorization 	&	Can exploit the improper/missing authorization  on SMO functions \cite{2}. & However, the attack range and possibilities are larger in Open RAN.		\\ \cline{2-3}
	&	DoS attacks 	&	Performs overload or flooding DoS attacks at SMO\cite{2}. &		\\ \cline{2-4}
	&	Orchestration associated security issues	&	Exploits weak orchestrator configuration, access control and isolation\cite{2}. &	Can be applied in V-RAN. However, the attack possibilities are higher in Open RAN.	\\ \hline
Hypervisor	&	VM/guest  OS manipulation	&	Exploits the security weaknesses in the guest OS to attack the hypervisor \cite{60,50,ferrag2018security}. &	The same type of attack can be applied in V-RAN. 	\\ \cline{2-3}
	&	Exhausting the hypervisor	&	Changes the configurations of compromised VNFs/CNFs to consume more resources and exhaust the hypervisor	\cite{60,50,ranaweera2021survey}. &	However, the attack range and possibilities are larger in Open RAN.	\\ \cline{2-3}
	&	Exceeding logs troubleshooting failure	&	Change the configurations of compromised VNFs/CNFs to generate excessive amounts of logs and exhaust the hypervisor\cite{60,50}. &		\\ \cline{2-3}
	&	Insider attacks	&	An insider who has access to the hypervisor misuses his privileges to perform an attack\cite{60,50}. &		\\ \hline
VM/CN	&	Misuse to attack others 	&	A VM/CN can be misused to attack another VM/CN, hypervisor/container engine, other hosts (memory, network, storage), etc.	\cite{2}&	The same type of attack can be applied in V-RAN. 	\\ \cline{2-3}
	&	Insecure run time configuration 	&	Insecure VM/CN run time configuration by the administrator can lower the security\cite{2}. & However, the attack range and possibities are larger in Open RAN.		\\ \cline{2-3}
	&	Compromise due to flaws 	&	VMs/CNs may be compromised due to flaws in the VNFs/CNFs they run\cite{2,tanakas2021novel,george2021preliminary}. &		\\ \cline{2-3}
	&	Attacker hack 	&	Hack into VM/CN retrieves the administrator privileges, resulting in obtaining all tenant's tokens and the administrator rights of the whole Open RAN system\cite{2}. &		\\ \cline{2-3}
	&	Malicious host 	&	The host OS has access to all data\cite{5,brandao2021hardening}. &		\\ \cline{2-3}
	&	Migration attack	&	During the VM/CN migration, a MITM attacker can modify arbitrary VM/CN OS or application states\cite{60,50}.	&		\\ \cline{2-3}
	&	VM rollback attack	&	An attacker uses and older snapshot of VM/CN to obtain access to the system\cite{60,50}.	&		\\ \cline{2-3}
	&	Scheduler Attacks	&	Misconfigures the hypervisor scheduler to allocate more resources to malicious VMs\cite{60,50}. &		\\ \cline{2-3}
	&	Eavesdropping on network traffic 	&	Eavesdrop on network traffic via a malicious VM/CN or hypervisor/container engine\cite{2}.	&		\\ \cline{2-3}
	&	Spoofing  on network traffic	&	Intercept and spoof on network traffic via VMs/CNs\cite{2}.	&		\\ \cline{2-3}
	&	Sharing hardware	&	Applications may share the same hardware resources in virtualization, which might be affected by vulnerabilities\cite{5}.	&		\\ \cline{2-3}
	&	Compromised security services	&	Compromises auxiliary/supporting network and security services\cite{2}. &		\\ \hline
 \end{tabular}
\end{center}
\end{table*}

The following components, Physical Network Function (PNF), Virtual Network Function (VNF), Cloud Network Function (CNF),  SMO, hypervisor, Virtual Machine/Container (VM/CN),  are involved in the virtualization process. Here, we discuss the security issues associated with each of these components.  Some of these attacks can be applied also in V-RAN and C-RAN\cite{hossain2019recent}. However, the attack range and impact of such attacks  larger in Open RAN. \color{black}The threats related to intelligence are summarized in Table \ref{VNrisks}.\color{black}


\begin{itemize}
\item \textbf{PNF related Attacks:}.
An attacker compromises a PNF to launch reverse attacks and other attacks against VNFs/CNFs.
A lack of security policies to protect mixed PNF-VNF/CNF deployments, resulting to insecure interfaces, could be exploited to perform attacks against VNFs/CNFs, potentially taking advantage of legacy security used by PNFs and not provided by the virtualization/containerization layer \cite{2}. \color{black} Apart from the security policies, service level agreements and service specifications are vital consensus for the PNF, CNF, and VNF entity deployment. As these entities are envisaged to launch security management entities as specified in \cite{benzaid2020ai}, the original consensus should not be altered for all the service level guarantees. \color{black} 

\item 	\textbf{VNF/CNF related Attacks:}
Despite VNF/CNF images are effectively static archive modules including all components used to run a given Open RAN VNF/CNF, modules within an image may have vulnerabilities, introducing malware, missing critical security updates or are outdated. These images are only collections of files packaged together. Therefore, malicious files can be included intentionally or inadvertently within them. 
In addition, VNF/CNF images may also have configuration defects, e.g. configuring a specific user with greater privileges than needed. 
This could all be used to attack other VMs/CNs or hosts within the environment.
An attacker can migrate a compromised VNF/CNF to a different location which has less security or privacy policies to gain additional access to the system. Since Open RAN uses different equipment with different vendors and different configurations, there can be less secure environments, which can  lead  to additional vulnerabilities  if  deployed in the same system \cite{2}.

Moreover, since many Open RAN VNFs/CNFs require secrets to enable authentication, access control and secure communication between components, these secrets are embedded directly into the image file system. In addition, the images often contain also sensitive components like an organization's proprietary software and administrator credentials. 
Anyone with access to the image (e.g. by means of insufficient authentication and authorization)  can easily parse it to extract these secrets, resulting in the compromise, stealing or damage of the contents on the images. 
As a result, it can lead to Intellectual Property (IP) loss and expose significant technical details about an Open RAN VNF/CNF image to an attacker. Even more critically, because registries of images are typically trusted as a source of valid, approved software, compromise of a registry can potentially lead to compromise of downstream VMs/CNs and hosts \cite{2}.

There is an increased risk of MITM attacks by intercepting network traffic intended for registries in order to steal developer or administrator credentials within that traffic. This can result in fraudulent or outdated images to orchestrators \cite{2}. \color{black} Further, typical VNF/CNF based security threats exist, as in: location shift attack where the adversary is capable of displacing the VNF to a domain inheriting a lesser level of security policy assignment with the intention of gaining access, or interoperability issues between the VNF/CNF developers or service providers that can be exploited by an attacker \cite{lal2017nfv,ranaweera2021survey}. \color{black}

\item 	\textbf{SMO related Attacks:}
\color{black} As the SMO is the key entity behind the holistic autonomic environment of the O-RAN, its security is extremely vital for the O-RAN performance and the individual subscriber security and privacy. \color{black} Improper or insufficient authentication or authorization of Open RAN external (e.g. AI/ML, Emotional Intelligence (EI), Human-Machine) or internal (e.g. over O1 or O2 interfaces, with Non-RT RIC) interfaces on SMO, allow access to the SMO and in particular the data stored on it. Besides disclosing Open RAN sensitive information, the attacker may also alter the Open RAN components  \cite{2}.

DoS attacks or increased traffic can cause overload situations and thus affects availability of the SMO data and functions. Further, an attacker may exploit weak orchestrator configuration, access control and isolation. A single orchestrator may run many different VMs/CNs, each managed by different teams, and with different sensitivity levels. If the access provided to users and groups is not conform their specific requirements, an attacker or careless user would be able to affect or subvert the operation of another VM/CN managed by the orchestrator. Malicious traffic from different VMs/CNs sharing the same virtual networks may be possible if VMs/CNs of different sensitivity levels are using the same virtual network with a poorly isolation of inter-VM/CN network traffic  \cite{2}.

\item 	\textbf{Hypervisor related Attacks:}
An attacker can exploit the security weaknesses in the guest OS to attack the hypervisor of the hosting OS. Examples of guest OS vulnerabilities are OS command injection, SQL injection, buffer overflow or missing authentication for critical functions  \cite{60,50, ferrag2018security}. \color{black} Privilege escalation is a common threat among hypervisor deployments that is also applicable in the context of the O-RAN. In this attack, any authorization violations are sought out by the perpetrator exploiting the infrastructure vulnerabilities formed through ill-maintenance or misconfigurations \cite{ranaweera2021survey}. The administrative capabilities granted to the adversary through this threat is devastating as it could range from a simple excessive allocation of resources to a complete deletion of xApps or rApps \cite{alnaim2019misuse,qiang2018privguard}. \color{black}

An attacker may also change the configurations of compromised VNFs/CNFs to consume high amounts of CPU, hard disk, and memory resources in order to exhaust the hypervisor. Another way to compromise the hypervisor is by generating an excessive amount of log entries such that it is infeasible or very difficult to  analyze the log files coming from other VNFs \cite{60,50}.

Finally, as the hypervisor provides its own security functions and Application Programming Interfaces (APIs) to the host system security functions, it is in full control of the security functionalities of the lower layers and thus needs to be fully trusted. When a malicious administrator has for instance  root access to the hypervisor and by using a search operation, the user identity (ID), passwords and Secure Shell Protocol (SSH) keys from the memory dump can be extracted, which in turn violates the user privacy and data confidentiality  \cite{60,50}. 

\begin{table*}
\caption{Overview of Global Open RAN risks}
\label{Globalrisks}
\footnotesize
\begin{center}
\begin{tabular}{|p{2cm} |  p{8.5cm}| p{5.5cm}| }
\hline
\rowcolor{gray!30}
  Threat & Description  & Specific to Open-RAN\\
\hline
\hline

Attack on digital economy	&	Network communications play an important role in the digital economy of a country and can cause huge damage in case of failure \cite{47, bugar2020techno} \color{black}, e.g. shutting down of smart cities, crashing of autonomous electrical vehicles or going dark of factories.  In particular, special attention should be given to avoid loss of trust with the users in case of such attacks as they might endanger the entire growth of the network \cite{spremic2018cyber}.\color{black}	&	These threats are very general and in particular related to attacks against the communication infrastructure. There are independent of the usage of Open RAN.	\\ 	
\hline
Espionage	&	Network communications can be abused to enable espionage	\cite{47,bederna2020cyber} \color{black}and there are currently no regulations for allowing or avoiding collaborations among different suppliers or actors in the network. Without being sure of good intentions of each of the involved entities offering the  equipment and software in the network, espionage at all levels, from the government to corporate, might be possible. As a consequence, it is extremely important that proper global ethical restraints are formulated on a global scale. \color{black}&	These threats are very general and in particular related to attacks against the communication infrastructure. There are independent of the usage of Open RAN. However, use of SW make Open RAN more vulnerable than RAN	\\ 	
\hline
Attacks on Critical infrastructure	&	Network communications play an important role in the operation, management and maintenance of critical infrastructures \cite{47}	\color{black}, like power grids, water supplies, manufacturing and transportation. Since the control of this infrastructure is handed over to the network operators, their responsibilities are of ultimate importance within the O-RAN domain. \color{black}&	These threats are very general and in particular related to attacks against the communication infrastructure. There are independent of the usage of Open RAN.	\\ 	
\hline
Violence against democracy	&	Besides espionage to dedicated people, also every other citizen can be envisaged.	\cite{47} \color{black}This might be a real threat to the democracy or freedom of speech in the world and it should thus be avoided in any case that one actor receives full control. \color{black}&	These threats are very general and in particular related to attacks against the communication infrastructure. There are independent of the usage of Open RAN.	\\ 	
\hline
Majority attacks and supply chain concerns	&	It is a danger if there  is a significant  involvement in Open RAN development from one country or one region, facilitating possibilities for  \cite{10,15} \color{black}any  type of attack. Special attention should be given that no new secret alliances are formed, and therefore a well balanced spread among the suppliers of O-RAN equipment and software is required. \color{black}&	These threats are very general and in particular related to attacks against the communication infrastructure. There are independent of the usage of Open RAN. However, use of SW make Open RAN more vulnerable than RAN	\\

\hline
	
 \end{tabular}
\end{center}
\end{table*}

\item 	\textbf{VM/CN related Attacks:}
VMs/CNs may be compromised due to flaws in the Open RAN VNFs/CNFs they run. For example, an Open RAN VNF/CNF may be vulnerable to cross-site scripting (SQL) injection \cite{tanakas2021novel} and buffer overflow vulnerabilities \cite{george2021preliminary}.

Insecure VM/CN runtime configuration by the administrator can lower the security of the Open RAN system. It may expose VMs/CNs and the hypervisor/container engine to increased risk from a compromised VM/CN. For example, it could be used to elevate privileges and attack VMs/CNs, the O-Cloud infrastructure/services, etc \cite{2}.

A compromised VM/CN will be able to alter that VM/CN in order to access other VMs/CNs, monitor VM/CN to VM/CN communications, attack the O-Cloud infrastructure/services, scan the network to which it is connected to in order to find other weaknesses to be exploited, etc. 
The container engine (in case of CN) or hypervisor (in case of VM) has access to all RAM memory, disk volumes mounted on the virtual machines and containers. This means that a malicious VM/CN or hypervisor/container engine can get access to all Open RAN network data processed in the workloads.
An attacker can launch a noisy neighbor attack against the shared O-Cloud infrastructure to cause the Open RAN system performance degradation and/or the services disruption by depriving the resources required by various Open RAN running functions \cite{2}.

An attacker hack into VM/CN is for instance possible if an attacker steals VMs/CNs private key from one VM/CN  and so reveals the administrator privileges. Next, all tenant's tokens and the administrator rights of the whole Open RAN system can be obtained, \cite{2}. 

From the side of the application, trust is required at all levels. In case the underlying host OS is malicious, access can be obtained to all data processed in the workloads, as in RAM memory and disk volumes. Techniques like secure enclaves \cite{brandao2021hardening} have the goal to provide a trusted environment. However, the application will be hardware-instance dependent \cite{5}.

If VM/CN migration is not secured or performed over a secure channel, a MITM attacker can modify arbitrary VM/CN OS or application states during the migration. An attacker may also use an older snapshot of VM/CN without the concern of the VM/CN owner to bypass the security system and obtain access to the system. This attack is possible after an already comprised hypervisor rollback to a previous snapshot. In the scheduler attack, the vulnerabilities  in  the  hypervisor's  scheduler are exploited to  acquire  system  resources  for  the malicious VM at the expense of a victim VM \cite{60,50}.

Furthermore, due to virtualization and cloud computing, different applications might use the same hardware resources. Isolation between these applications are only at software level and not at the level of hardware. As a consequence, hardware related vulnerabilities like the recently discovered Meltdown and Spectre attacks (https://meltdownattack. com/) can have a larger attack range \cite{5}.

Finally, besides the main functionality of the VNF/CNF itself, the administrators may also decide to deploy additional network services on their VMs/CNs in order to do extra monitoring, remote configuration, remote access to other services such as SSH, etc. If these additional network services are directly accessible over the Internet or from another administrator, new entry points for attackers are created and if access is obtained to the VM/CN, more extra attacks become possible \cite{2}.
\end{itemize}

\subsection{Global}
Offering the highest level of security on the network is important for a nation. We here distinguish five major types of attacks or risks that need to be taken into account \cite{47}.
\color{black}The related threats can be found in Table \ref{Globalrisks}.\color{black}


\begin{itemize}
\item \textbf{Attack on digital economy:} Since  5G is fully integrated in the digital economy, it can result in potential life or death consequences. For instance, currently a lot of data is sent from our mobile devices, smart homes, electrical cars via a network consisting of devices, which are remotely controlled and updated and thus present a potential attack vector. The possibility of a smart city shutting down, autonomous vehicles crashing, or factories going dark due to a cyber attack are frightening situations; \color{black}that would eventually result in a major economic collapse. At this pivotal point in modern civilization where the global economic platforms are shifting to a holistic digital platform, a successful threat might endanger the entire growth of 5G and its predecessors through loss of trust from the subscribers. Thus, it is imperative to investigate the scope of such threat vectors that target economic platforms. It is evident the prescribed scope is reaching beyond the means of typical phishing, or identity thefts \cite{spremic2018cyber}. Moreover, the impending launching of Metaverse and its significance for O-RAN existence is further confirming the required focus on the robustness of digital economic platforms, as Metaverse is introducing a virtual serviceable platform built on top of monetary transactions \cite{chang20226g}. \color{black}

\item \textbf{Espionage:} There are currently no regulations for avoiding the collaboration between an Open RAN equipment manufacturer and an external party, like for instance a security agency of a certain country. Therefore, without no guarantee of good intentions of the  equipment and software providers, possibilities for spying should be considered as viable. \color{black} Such acts of espionage can be perpetrated by targeting corporate to government institutions. The flexibility offered through O-RAN standardization might be exploited, and privacy violations become the least of concerns for network operators. The AI-based decision making the backbone of the O-RAN architecture is inviting instilling of botnet-type autonomous constructs that entail a sophisticated cyber intrusion; where prevention is quite arduous \cite{bederna2020cyber}. Therefore, proper ethical restraints should be drawn in a global scale to prevent such acts of espionage, while monitoring to detect such acts are equally pertinent. \color{black}

\item \textbf{Attacks on critical infrastructure:} Critical infrastructure typically consists of the management of  power grids, water supplies, manufacturing, and transportation infrastructure.
More and more, 5G is used as the backbone of communication in these infrastructures. Therefore, a dedicated cyber attack disrupting this critical infrastructure would have a devastating impact on the people dependent on this. \color{black} Conversely, the control of the critical infrastructure is handed over to the 5G network operators. Therefore, their responsibility is ever so critical and honourable. Since the government level acts for blocking critical infrastructure to deliver threats in the geo-political arena are not rare occurrences, O-RANs dependence on the same network operators is raising concerns in the global shared resource market. Thus, the responsibilities of the network operators become extremely important within the O-RAN domain. \color{black}

\item \textbf{Violence against democracy:} If an actor receives the power to perform the role of big brother in all communication, there is a real threat for democracy and freedom of speech. \color{black} As all the means of global economic infrastructure are envisaged to be shifted to a digital environment backed by the 5G enabled networks, democracy becomes merely a concept without any context or standing in case of a total takeover. The ideals that made O-RAN more efficient and flexible might lead to the downfall of modern democracy and its stance on the global scale.\color{black}

\item \textbf{Majority attacks and supply chain concerns:} As mentioned in \cite{10,15}, the Open RAN Alliance currently includes a wide range of high security risk companies. If the efforts in the development and standardization process for Open RAN is dominated by partners belonging to one country or even one region, it can cause an imbalance resulting in a new alliance that will still enable espionage possibilities and disrupt the intended openness. \color{black} As supply chains formed through globalization are relying on online trading and financial platforms for international transactions, the responsibility of the O-RAN stakeholders is ever so vital in facilitating the required digital infrastructure. It is obvious that Blockchain serves as an appropriate solution to secure such a financial infrastructure. The majority of attacks or 51\% attacks are however, proving to be realistic, where an attacker is capable of withdrawing the payment after the merchant has sent the product \cite{dey2018securing}. This threat is intimidating the credibility of the Blockchain networks by enabling plausible deniability - which is one of its foremost purposes for the emergence of Blockchain. \color{black}
\end{itemize}

\section{Open RAN Best Security Practices} 
\label{sec:solution}

We discuss the best security practices for Open RAN in this section. As Open RAN is a derivative of the conventional C-RAN, it will inherit many threats and vulnerabilities of C-RAN. Therefore, a number of C-RAN security solutions can be adopted by Open RAN without any significant modifications. For example, the existing security solutions to prevent primary user emulation attacks (PUEA) can be adopted for Open RAN~\cite{CRAN_security}. In~\cite{PUEA1}, the authors discussed cryptographic and wireless link signatures to distinguish between a legitimate user from an attacker. A helper node is proposed that is placed around a primary user. The helper node acts as a bridge between the primary and secondary users by sending authentic link signatures to the secondary nodes. The authors also proposed a corresponding physical layer authentication algorithm in~\cite{PUEA1}. There are other security mechanisms against PUEA based on the received signal strength. 
In~\cite{PUEA2}, the authors proposed naive detection and variance detection methods against PUEA. The authors modeled advanced strategies of PUEA where both the legitimate user and the attacker can exploit estimation techniques and learning algorithms. The variance detection attack is effective against PUEA for a time-invariant channel. 


The most widely researched security threat in the medium access control layer is the spectrum sensing data falsification (SSDF) attack where an active attacker transmits error observation to disrupt collaborative spectrum sensing and resource allocation~\cite{CRAN_security}. A joint spectrum sensing and resource allocation scheme is proposed in~\cite{SSDF} to combat the SSDF attack. The problem is formulated as a weighted-proportional-fairness-based optimization problem with an additional constraint of the primary user being sufficiently protected. The authors decomposed the problem into two subproblems which are a resource allocation problem and a cooperative secondary user decision problem. The key idea of the scheme lies in improving the secondary users' sensing reliability and preventing the secondary user from acting maliciously. The computer simulations showed that the proposed scheme deals with the SSDF attack in the co-operative sensing process to improve system robustness. 

As the Open RAN systems adopt cloud computing unlike conventional RAN systems, the security solutions for cloud computing are also relevant for Open RAN. In~\cite{cloud_computing_sec}, the authors identified security and privacy vulnerabilities of cloud computing that can be exploited by an adversary for various attacks. The authors presented basic requirements to build a secure cloud system by addressing three main challenges, namely, outsourcing, multi-tenancy, massive data and intense computation. To address the outsourcing challenge, the cloud provider needs to provide secure and trustworthy data storage. The outsourced data also needs to be verifiable by the customer. For multi-tenancy, the cloud platform needs to securely perform resource allocation in the virtualized environment. Finally, massive data sets need to be broken down into small sets to accelerate the processing. The solutions to PUEA, SSDF and cloud computing vulnerabilities are applicable to both C-RAN and Open RAN. We invite interested readers to go through~\cite{CRAN_security} for more discussion on the C-RAN security vulnerabilities and solutions.  

We identify three key components to resolve security vulnerabilities that are exclusive to Open RAN. The first component to enhance Open RAN security is blockchain-based mutual authentication. As O-RAN promotes openness between a pool of untrustworthy O-RU and O-DUs unlike C-RAN, the blockchain can be a very important and unique tool to establish trust between them and enable a safe communication mechanism. The second key component is the physical layer itself. The difference from other RAN systems is the operators have the option to select O-RUs from different competing vendors in Open RAN technology. Thus, the O-RUs can be installed at any moment with the desired number of antennas, front-end processing, beamforming algorithms to enhance the security of the Open RAN. We also discuss RF-fingerprinting techniques which can be crucial to identifying rogue RUs trying to connect to the system. The third key component is AI algorithms. As Open RAN provides more interfaces to enable an intelligent RAN system, we discuss a few examples of AI-enabled enhanced security in Open RAN. Table~\ref{security_sol_table} summarizes the key solutions to threats and vulnerabilities related to Open RAN. Finally, we present a subsection regarding the common mistakes in Open RAN design, their consequences and mitigation.

\begin{table*}
\caption{Security Solutions to Open RAN risks}
\label{security_sol_table}
\footnotesize
\begin{center}
\begin{tabular}{|p{3cm} |  p{8cm}| p{5cm}| }
\hline
\rowcolor{gray!30}
  Threats and vulnerabilities & Solutions & Specific to Open RAN\\
\hline
\hline

Primary
user emulation attack	&	 A helper node can be used proposed which acts as a bridge between
the primary and secondary user~\cite{PUEA1} or a variance detection method is adopted based on received signal strength~\cite{PUEA2}	&	No, applies to both C-RAN and Open RAN	\\ 	
\hline
Spectrum sensing data falsification	& A weighted-proportional-fairness-
based optimization problem can be formulated with an additional constraint of
primary user being sufficiently protected~\cite{SSDF} &	No, applies to both C-RAN and Open RAN	\\ 	
\hline
Cloud computing vulnerabilities	&	A secure cloud system needs to address three main challenges, namely, outsourcing,
multi-tenancy, massive data and intense computation~\cite{cloud_computing_sec}&		No, applies to both C-RAN and Open RAN	\\ 	
\hline
Untrustworthy O-DUs and O-RUs	& A blockchain-enabled RAN framework can establish trust between untrustworthy O-DUs and O-RUs through a smart contract which is verified by third party miners~\cite{BRAN} &		Yes, only relevant to Open RAN due to its inherent openness	\\ 	
\hline
Privacy concerns in P2P communication	&	A distributed identity authentication through blockchain can enable privacy preserving P2P communication without the involvement of certificate authority or public key infrastructure~\cite{Blockchain_privacy_ORAN}&		Yes, because Open RAN architecture enables different P2P communications such as D2D, M2M, etc\\ 	
\hline
Rogue O-RU	& With prior knowledge of the transient and steady-state response of power amplifiers and other RF circuitry, RF fingerprinting techniques can determine whether an O-RU is rogue or benign~\cite{RF_fingerprint_review}&		Yes, this threat is only relevant to Open RAN as it allows to integrate O-RUs from different vendors	\\ 	
\hline
Eavesdropping	& Increasing the number of antennas and corresponding digital front-end processing with beamforming capabilities can diminish the threat of passive eavesdropping~\cite{massive_mimo_security} and increase the probability of active attack detection~\cite{massive_mimo_security2}	 &	Yes, the solution is O-RAN specific because the operator has the freedom to choose suitable O-RUs and O-DUs from different vendors in O-RAN	\\ 
\hline
Conventional security framework	&	Open RAN enables intelligent zero-trust security framework upon which advanced AI algorithms can be developed to provide security in untrusted networks~\cite{ramezanpour2021intelligent}  &	Yes, the proposed framework in~\cite{ramezanpour2021intelligent} adopts service based design by leveraging Open RAN architecture to ensure ease of integration	\\ 
\hline
DDoS attacks	&	Open RAN systems can employ machine learning algorithms that are
trained to protect the network from DDoS attacks with very high accuracy~\cite{doshi_ddos,sharafaldin_ddos} &	No, the machine learning solutions to detect DDoS attacks can be used by any RAN system	\\

\hline
	
 \end{tabular}
\end{center}
\end{table*}

\subsection{Blockchain-enabled Open RAN} \label{BLRAN}
Blockchain or distributed ledger technology (DLT) is a distributed database for exchanging identities of users and storing records of all user identities that are linked together using cryptography \cite{lipton2021blockchain}. In other words, it is a chain of interconnected information blocks that creates a public ledger for recording a list of transactions. Blockchain is popular for its crucial role in modern cryptocurrency systems to provide a secure and decentralized record of transactions \cite{ur2019trust}. Blockchain or DLT has also emerged as a tool for designing a self-organized and secure radio access network (RAN). Blockchain-enabled identity management and authentication can lower the cost and aid the core network to provide more secure and user-oriented service in an era of open and distributed RAN deployments  \cite{9672678,9148820}. In~\cite{BRAN}, a RAN framework has been presented by leveraging the principles of blockchain. This framework proposes that the UEs and APs in the network agree about payments or spectrum assets based on a contract. The terms of this agreement are recorded by a smart contract, authorized by client signatures. Afterward, the contract is verified by the miners to determine whether the UEs have sufficient credit balance or the APs have sufficient spectrum assets. The verified contracts are aggregated to a block, which is added to the existing blockchain. In this process, a UE will be granted limited-time access to the spectrum assets while an AP can receive the payment automatically. As a result of enforcing the rights of relevant parties by means of a smart contract, trust has been established between initially untrustworthy UEs and APs. The application of blockchains and smart contracts for RAN can be further extended to cooperative communication, mobile ad-hoc networks and privacy-preserving communication systems. 

As the RAN technology is moving towards more open, intelligent, virtualized and interoperable networks in the form of Open RAN, it will be crucial to develop trust between a pool of O-RU and O-DU vendors. A blockchain-enabled smart contract establishes trust between these vendors and provides a mechanism to constantly monitor the development of the system by independent third parties. One such example is a blockchain-enabled privacy-preserving point-to-point (P2P) communication in Open RAN. Due to the advent of distributed and decentralized functionality of Open RAN, different P2P communication in mobile networks such as device-to-device (D2D) or machine-to-machine (M2M) will be benefited. However, the fundamental security flaws for P2P communication in a mobile network will remain in an Open RAN. The P2P has limitations in global peer discovery and routing without third party assistance and thus, the coverage is a bottleneck. In the centralized architecture, the UE is restricted to communicate directly with other users. Two users under the same BS cannot be directly connected to each other without the involvement of the core network. The reason is several key functionalities such as identity authentication, routing, etc, are only done at the core network in the state-of-the-art mobile networks. The distributed identity authentication issues in the current architecture can be addressed by blockchain due to its decentralized nature. The identity authentication can be performed locally at the RAN with a global identities record. In this way, two users under the same RAN unit can communicate with each other directly without accessing the core network. 

In~\cite{Blockchain_privacy_ORAN}, a blockchain-enabled mutual authentication architecture is presented for identity management in Open RAN that does not require a third-party Certificate Authority (CA)
or Public Key Infrastructure (PKI). The authors proposed a blockchain address (BC ADD) as a global identity for all UEs within a RAN, where all users generate their own address by hashing their public keys. These newly generated addresses are recorded by the ledger records and used as anonymous identities locally or globally. The relationship between the public key and BC ADD is strictly one-directional. As a result, it is difficult to fake a BC ADD when the pubic key is known, or vice versa. The authors compared the performance of their proposed mutual authentication method based on blockchain with   Internet Key Exchange
version 2 (IKEv2) and Transport Layer Security (TLS) from signaling, communication and
computation perspectives. The authors noted blockchain based scheme only requires 2 signals while IKEv2 and TLS 1.3 require 4 and 9 signals, respectively. The authors also demonstrated that the blockchain method requires significantly less number of bytes for finite-field and elliptic curve cryptography (ECC) for communication and computations. For communication, the blockchain method requires 1060 and 356 bytes for finite-field and ECC, respectively. The IKEv2 requires 3820 and 3110 bytes for finite-field and ECC, respectively, for the same functionality. Similarly, the number of bytes required for both finite-field and ECC is significantly higher for computation in IKEv2. 

In~\cite{velliangiri2021blockchain}, the authors presented a privacy preserving framework of blockchain enabled RAN for increased efficiency and enhanced security. The authors simulated their system model in Hyperledger Fabric 1.2 based simulator. The simulation shows that the blockchain enabled RAN achieves higher throughput and lower resource consumption compared to conventional RANs. As Open RAN promotes open source software development for the base stations, it is imperative to develop a distributed security mechanism with {\it many eyes} to observe the changes in the Open RAN operation. Blockchain can provide an ideal framework to support RAN elements from different suppliers in a secured and organized manner. \textcolor{black}{We believe blockchain can be a key element in future ORAN systems for authentication and identity management. However, several challenges remain to integrate blockchain technology into wireless networks. For power-limited node devices and cost-sensitive transmission networks, implementing blockchain based mutual authentication can be challenging. In addition, latency can be a critical issue of blockchains for delay-sensitive scenarios in a wireless network. Despite these challenges, blockchain can be an important component of the Open RAN systems as they suffer from more security challenges than traditional RANs due to their openness by design.}

\subsection{Leveraging physical layer to enhance Open RAN security}
From an O-DU's perspective, it is crucial to differentiate a legitimate O-RU from a masquerading O-RU in the Open RAN systems. The transmitter of the O-RU consists of radio frequency (RF) modules such as digital-to-analog (DAC) converters, power amplifier (PA), analog band-pass filters, frequency mixers, etc. Despite decades of research and development effort by the microwave circuits community, longstanding imperfections still exist in the RF transmitter chain. These imperfections cannot be altered or corrected without significant effort and thus, can be exploited as radiometric signatures of different O-RUs. In addition to conscious design decisions, these imperfections can stem from uncontrollable factors in the manufacturing process such as differences in the semiconductor doping industry. As a result, different O-RUs can have very different flatness and ripples in the RF spectrum, differences in rejection and transition bands, mismatch in the I/Q phase, DC offset or gain imbalance, etc. The idea of using RF fingerprints to identify devices through such intrinsic features of the RF stages has been widely explored by the microwave circuit community. We believe such RF fingerprinting can also be used as the first line of defense to detect a masquerading O-RU. 

A review of RF fingerprinting techniques has been presented in~\cite{RF_fingerprint_review}. The authors described state-of-the-art techniques for RF fingerprinting based on transient response. These methods utilize the transition from the turn-off to the turn-on of a power amplifier that occurs before the start-up of a radio unit. The transient response of every power amplifier is unique and thus, can be used for wireless device identification. However, this method is effective when the transient is accurately known, i.e. the exact beginning and the exact end. The authors discussed several methods for detecting the start point of the transient, such as Bayesian step change detection, Bayesian ramp-up change detection, phase detection, mean change point detection, etc. In~\cite{brik2008wireless}, the authors proposed an identification system based on the steady-state response of the hardware. The system is called passive radiometric device
identification system (PARADIS) and uses five features: frequency error, correlation, I/Q offset, magnitude errors and phase errors to identify a device. As detecting the transient response requires a very high sample rate, which is infeasible in many applications, the steady-state response is frequently used for RF fingerprinting.

In~\cite{PA_identification}, a model based approach is presented for the identification of wireless users via power amplifier imperfections. The authors exploited the differences in non-linearities of I/O characteristics of a power amplifier modeled with the Volterra series. The authors proposed a generalized likelihood ratio test (GLRT) and a classical likelihood ratio test to identify the legitimate user. A symbol based statistical RF fingerprinting technique for fake base station identification is presented in~\cite{RF_fingerprinting_stat}. The authors present a scheme to detect unique non-linearities based on hardware impairments of the transmitter. The proposed scheme is based on the assumption that a fake base station tends to violate the spectral mark and introduces large amplitude and phase errors compared to a legitimate base station. The RF fingerprinting can be an ideal mechanism to verify that an O-RU is secure enough to be connected to the O-DU of the O-RAN.   

A massive multiple-input multiple-output (MIMO) O-RU can also improve the security in an Open RAN system. A massive MIMO system equips the base station with a large number of antenna elements which can serve a large number of user terminals in the same frequency band~\cite{larsson2014massive}. It should be noted that the antennas reside in the O-RU of Open RAN while the baseband layer processing is performed in the O-DU. The number of layers in baseband is typically 16 or less in a 5G base station. For an $N$-layer O-DU, the O-RU must support at least a number of $N$ antennas and RF-front end circuitry. If the number of antennas in O-RU is significantly higher (e.g. 8-10 times) than $N$, the Open RAN system can be considered a massive MIMO system. Every layer of baseband data in the O-DU can exploit the higher number of antennas in O-RU with beamforming techniques. The base station can direct its baseband data in a specific direction by constructively adding multiple antenna streams and improving the signal quality. Due to their beamforming capability, the massive MIMO systems are more secure than the small-scale MIMO systems. It is possible to direct a narrow beam toward a legitimate user in a massive MIMO beamforming system. If an eavesdropper is not in the vicinity of the legitimate user, the received signal power of the eavesdropper is significantly diminished while the received power of the legitimate user increases manifold.

In~\cite{massive_mimo_security}, the authors presented analytical results that showed a passive eavesdropper has a negligible effect on the secrecy capacity in a massive MIMO system. Their simulation shows that a passive eavesdropper's capacity remains the same with an increasing number of antennas. However, the legitimate user's capacity increases greatly for a large number of antennas. For a small-scale MIMO system with 2-8 antennas, the legitimate user's secrecy capacity is about half of the channel capacity. When the number of antennas is 100, the secrecy capacity reaches about 85 percent of the channel capacity. The primary reason for the resilience of a massive MIMO system against a passive eavesdropper is based on the assumption that the uplink channel estimation is independent of the eavesdropper's channel. However, an active eavesdropper can transmit pilot signals to the base station to influence the base station's transmit beamforming design. In such a scenario, the physical layer security of a massive MIMO system is compromised and the achievable secrecy rate vanishes with the increasing power of the eavesdropper's pilot signal. However, the probability of detecting an attack increases with increasing eavesdropper's signal power~\cite{massive_mimo_security2}. Two active eavesdropper detection methods have been proposed in~\cite{massive_mimo_security}. The first scheme is based on random quadrature phase-shift keying (QPSK) pilot transmission by the legitimate user. The idea is that the phase of two legitimate pilot signals converges to valid PSK symbols as the number of antennas is large. In the second scheme, the beamformer is constructed in such a way that the received signal at the legitimate user is equal to an agreed value. These two detection schemes are only effective due to the large number of antennas in a massive MIMO system. Due to their centralized structure, the current base stations typically employ a fixed number of antennas and baseband layers. In the Open RAN design paradigm, the operators can select an O-RU with a higher number of antenna chains and thus, with a capability to beamform and enhance security. \textcolor{black}{We believe the ability to select O-RUs with the desired configuration will be crucial to improving the overall security of an Open RAN system. The most popular physical layer candidates for future wireless standards such as cell-free MIMO or reflective intelligent surface (RIS) can utilize a high number of antennas. Thus, applying specific physical layer configurations is a viable solution to combat security threats such as eavesdropping in an Open RAN system. Identifying rogue O-RU will be crucial for Open RANs to succeed and replace the conventional RANs. We believe RF fingerprinting could be the first line of defense against a rogue O-RU that is trying to connect to the network.}

\subsection{AI enabled Open RAN Security}
Since the introduction of deep learning by Hinton {\it et al.}, there has been a reinvigorating interest in AI applications in the wireless communication research community. The ML based solutions have also been popular in the network security research community. Despite some security concerns associated with AI based solutions as mentioned before, the AI automated security solutions will represent an essential key element of future wireless networks. The application of AI is so crucial that entire security frameworks have been proposed to utilize the AI algorithms. In~\cite{ramezanpour2021intelligent}, an architectural concept design of an intelligent zero trust architecture upon which advanced AI algorithms can be developed is proposed in order to provide security in untrusted networks. This framework adopts a service-based design by leveraging Open RAN architecture to ensure ease of integration. The three main components of zero-trust architecture in Open RAN are intelligent agent or portal (IGP), intelligent network security state analysis (INSSA), and intelligent policy engine (IPE). 

The IGP employs a reinforcement learning approach to analyze the incoming traffic, provides an initial risk assessment, and a model for their security posture. The reinforcement learning model used by multiple IGPs can be a common model that is trained in the federated learning approach. By utilizing federated learning, a more comprehensive model of the local environment is trained by different subjects. The second component INSSA provides a dynamic risk assessment for every access request. The authors proposed a graph neural network to model the state of Open RAN. The neural network models the communication patterns of the Open RAN with the goal to assign risk scores in such a way that the overall security metric is maximized while granting access. The final component of the zero trust architecture in Open RAN is the IPE which takes the final decision to grant access. The IPE is based on a neural network called long short-term memory (LSTM) to evaluate the risk of granting access based on reports from IGPs and INSSA. After making a decision, the IPE monitors the security state of the session. The IGP, the INSSA, and the IPE work together to provide a cohesive framework for zero trust in Open RAN.    

Conventional hardware dependent security such as firewalls or deep hardware inspections might not be the ideal solution for a dynamic and open environment of Open RAN. Therefore, it would be crucial to develop automated mechanisms for intrusion detection, attack response and mitigation. An Open RAN system can employ an ML mechanism that is trained to protect the network from DDoS attacks. A plethora of ML based mechanisms for DDoS detection can be found in the literature. Five classification methods, including $K$-nearest neighbors (KNN), Decision Tree (DT), Random Forest
(RF), Support Vector Machine with linear kernel (L-SVM), and Neural Network (NN) have been studied for intrusion detection in~\cite{doshi_ddos}. The authors used a limited set of features to enable real-time classification and middlebox deployment. The authors found that all five methods were able to detect DDoS attacks with a high level of accuracy. However, the authors considered only three types of DDoS attacks. A total of 13 different DDoS attacks were considered in~\cite{sharafaldin_ddos}. The accuracy of the ML algorithms decreased significantly for this scenario. In addition, both works of~\cite{doshi_ddos} and~\cite{sharafaldin_ddos} used supervised learning which requires labeled data. 
Such labeled data can be challenging to obtain and thus, the application of supervised learning is not always realistic. 

In~\cite{ddos_unsupervised}, the authors discussed network intrusion detection systems using different autoencoder architectures. Autoencoders are a type of artificial neural network based on unsupervised learning that aims to reconstruct its original input vectors. The proposed intrusion detection autoencoder develops a threshold heuristic of the reconstruction error which represents the proportion of abnormality in training data. The authors considered four types of autoencoders namely basic autoencoder, stacked autoencoder, denoising autoencoder, and variational autoencoder. The results showed that stacked and variational autoencoder perform better than the rest.
A time-based anomaly detection system, named Chronos, is presented in~\cite{salahuddin2021chronos}. Chronos is an autoencoder that utilizes time-based features to detect anomalous DDoS traffic. This method extracts statistical information from time-based features for each small set of packets collected during a time window. The efficacy of Chronos was evaluated by performing extensive evaluations on the CICDDoS2019 dataset. The authors also evaluated the impact of different window sizes to detect DDoS attacks. Chronos achieves an accuracy of over 99\% for most attacks and greater than 95.86\% for all attacks.

The application of AI for Open RAN security is not limited to intrusion detection. AI is an effective tool to identify devices based on RF fingerprinting. Contrary to the hand-engineered approaches, the ML approaches are able to rapidly identify a rogue O-RU before sharing any network information. In~\cite{rf_fingerprinting_ML}, the authors presented a convolutional neural network with a triple loss for RF fingerprinting. The authors demonstrated the feasibility of the proposed scheme over the experimental POWDER platform in Salt Lake City, Utah, USA. The proposed method achieves a 99.86\% detection accuracy for different training and testing days on real-world datasets. 

In~\cite{rf_fingerprinting_ML2}, the authors studied four ML techniques to identify RF devices in the time domain. These four schemes are deep neural networks, convolutional neural networks, support vector machines, and
multi-stage training using accelerated Levenberg-Marquardt. The authors examined data originating from 12 different transmitters. The accelerated Levenberg-Marquardt based training method achieved 100\% accuracy and outperformed state-of-the-art ML methods. A massive experimental study of deep learning for RF fingerprinting has been presented in~\cite{rf_fingerprinting_ML3}. The authors analyzed 400\,GB of I/Q data transmitted by 10,000 radios. The authors chose convolutional neural networks because of their ability to interpret features better than conventional ML techniques. This work demonstrated that the proposed solution can handle different channel conditions and signal-to-noise ratios, and is scalable to very large populations.
\textcolor{black}{Its almost certain that AI will play an integral role in the security of Open RAN systems. However, ML techniques themselves are vulnerable to security threats. The effectiveness of ML algorithms greatly depends on the quality of training data sets. An adversary can also send false data during the training process of the system. Therefore, robust ML algorithms that can tolerate malicious inputs need to be adopted. In addition, stability training can be adopted so that the ML schemes do not deteriorate for different and independent data sets.}

\subsection{General mistakes, consequences and mitigation}

Almost all of the technology related attacks mentioned in Section \ref{techrisks} are caused because of some general design errors. Table \ref{br} summarizes these major errors, together with their consequences and potential mitigation measures\footnote{Note that these mitigation measures are closely related to the process related risks, which can in fact also be considered as guidelines (see Table \ref{prrisks})}.

\begin{table*}

\caption{Overview of most common errors, consequences and mitigation measurements}
\footnotesize
\begin{center}
\begin{tabular}{ |p{2.5cm} | p{6cm} | p{7cm}| }

\hline
\rowcolor{gray!30}
General error & Consequences & Risk mitigation  \\
\hline
Insecure design of Open RAN interfaces  & \color{black} Novel design strategies can be left with critical flaws due to the open and flexible approach of O-RAN \cite{mimran2022evaluating}. Malware injection resulting in DoS attacks to retrieval of sensitive information via unauthenticated/unauthorized access \cite{burakovsky2022imperative}.

& \color{black}  Define security standards and protocols, as in Media Access Control Security (MACsec) for Open RAN devices and interfaces \cite{dik2021transport}. Issue security certificates via standardized bodies. Train people to apply the defined standards and processes. Monitor network traffic for suspicious activity on all levels. Add firewalls and rate limiters to act appropriately. Proper Access control mechanism should be deployed as SbD \cite{klement2022open}. 
\\
\hline

Software flaws  & Software for firewall protection can result to failures. Exploitation of  buffer overflows results in the execution of arbitrary commands with devastating consequences.    
& \color{black} Be careful with open source software and keep software always up-to-date. Invest in security training for employees. Purchase software from trusted suppliers and use third party certificate authorities. Follow SDS approach for automating flaw detection \cite{harer2018automated,chernis2018machine}.
\\
\hline
Insufficient protection of security event log files.
&  Security restoration delays, wrong audits and threats persistence.
& \color{black} Automate the log monitoring process and add rate limits. Define log management clearly in the standard. SbD approach for automating log maintenance while introducing anomalous log detection using ML \cite{yadav2020survey,klement2022open}. \\
\hline
Insufficient protection of data storage 
& Attacks against privacy, including data tampering, information disclosure, elevation of privilege, etc.  
&
\color{black} MACsec protocol suite can be followed for the fronthaul, while SDS based approach can be deployed for the network and application layer network automation \cite{blanc2018towards,klement2022open}. Follow data poisoning prevention methods \cite{desai2020cache}. 
\\
\hline
Compromise of integrity and availability 
&
DoS attacks. See also consequences of first error in case of improper authorization and authentication
& \color{black} A proper autonomous authentication and authorization mechanism is required to protect integrity and ensure availability \cite{ramezanpour2021intelligent}. \\
\hline
Physical access 
& 
Retrieval of stored private keys, certificates, user plane data, control plane data and management data in cleartext.
Modification of  Open RAN components settings and configurations in order to disable security features and allow eavesdropping or wiretapping on various planes, creates  performance issues  \cite{5}.
&Define security standards for physical security. Ensure that all stakeholders in the Open RAN system are identified, authenticated and trusted \cite{9}. 
\\
   \hline
 \end{tabular}
\end{center}
\label{br}
\end{table*}

The following six general mistakes are identified.
\begin{itemize}
\item  \textit{\textbf{The hardware-software Open RAN system suffers from insecure design:}}

\color{black} The open fronthaul and its interfaces, xApps based radio resource management, Decoupled hardware, open management interfaces, and open source deployments are some insecure design strategies that require more investigation to determine remedial possibilities \cite{mimran2022evaluating}.   \color{black} This includes misconfigured or poorly configured Open RAN interfaces due to the outdated components or improperly configured permissions, insufficient/improper mechanisms for authentication, encryption and authorization in different hardware-software components of the Open RAN system.

This type of weakness would allow attackers to inject malware in order to manipulate and harm the Open RAN components, which may result in a variety of consequences going from launching DoS attacks to retrieval of sensitive information including unprotected private keys. As a consequence, the attacker gets unauthenticated/unauthorized access to Open RAN components via the different Open RAN interfaces. 

In order to offer protection against this, security standards (e.g. Media Access Control Security - MACsec) for Open RAN devices and interfaces should be clearly defined and people should be trained in order to apply these standards and processes \cite{dik2021transport}. In particular, special attention should be given to all access control mechanisms  at every access point. \color{black} A Security by Design (SbD) approach might be suitable for automating the authorization framework \cite{klement2022open}. \color{black} Security certifications should be issued via trusted security standardization bodies. It is also essential to monitor network traffic for suspicious activity on all levels and to add firewalls and rate limiters to act appropriately. 

 \item \textit{\textbf{Software flaws:}} \color{black} The tendency to utilize open source solutions for virtualization deployments as well as for the RU, DU, and CU based deployments is a way to assert the required interoperability within a multi-vendor O-RAN; At the same time exposes the software specifications and configurations to the general public \cite{mimran2022evaluating}. This opportunity allowing the assimilation of the software standards would benefit either negatively or positively depending on the capability of the attacker or the defender. \color{black} Software flaws are present in the different components of the Open RAN system, which are not notified and mitigated in time.
 
 In particular, software providing network functionalities play an important role in firewall protection. If vulnerabilities like buffer overflows are exploited, arbitrary commands can be executed with devastating consequences.
 
 As Open RAN is built with open source software, it is important to keep the software always up-to-date and to make sure that they are developed by trusted suppliers, which use third party certificate authorities. Security training for employees is required such that software is developed with the highest standards. \color{black} Further, Software Defined Security (SDS) concept can be extended to the application layer from the network layer for automating the security functions and detecting software based flaws through machine learning means \cite{chernis2018machine, harer2018automated}.\color{black}
 
 \item \textit{\textbf{Security event log files, generated by the different Open RAN components, are not sufficiently or improperly protected:}} \color{black} Security has not been recognized as a function by the O-RAN setup to maintain proper and organized logs for event recording.\color{black} For instance, they lack information on host name, IP or MAC address, correct timing of incidents, etc.
 
 Compromise or incomplete logs can result in security restoration delays, wrong audits and threats persistence.
 
 The standard should clearly define how to manage the log monitoring process and  rate limits should be added. \color{black} SbD approaches can be employed for centralized log auditing, while an automated anomalous detection method for security logs can be an efficient directive to overcome this pitfall \cite{yadav2020survey,klement2022open}. \color{black}
 
 \item \textit{\textbf{Sensitive data, stored, processed and transferred among the different Open RAN components, are not secured according to industry best practices:}} For instance, appropriate encryption and integrity protection mechanisms are missing, inappropriate access control, lack of traceability of the access in the audits, etc.
 
 This weakness will mostly result in attacks against privacy, including data tampering, information disclosure, elevation of privilege, etc.  
 
\color{black} Security standards for authentication and end-to-end encryption (MACsec) following the latest developments should be defined \cite{dik2021transport}. Security certifications should be issued via security standardization bodies. People should also be trained to apply the defined standards and processes. Finally, data poisoning prevention algorithms should be deployed \cite{desai2020cache}. \color{black}
 
 \item \textit{\textbf{Integrity and availability of Open RAN components can be compromised:}} Integrity and availability of Open RAN components can be compromised due to overload situations caused by DoS attacks or increased traffic and where the Open RAN components do not possess the required functionalities to deal with it. 
 
 A direct consequence is of course a DoS attack. However, other possibilities may appear in case of insufficient or improper configuration (see first error). For instance, an attacker might be able to boot Open RAN components from unauthorized memory devices and thus instill a selected malware to a xApp or any other operational entity \cite{2}.
 
 Similar counter measures as in the previous mistake should be taken. 
 
 \item \textit{\textbf{Several possibilities allow physical access to different components in the Open RAN system:}} First, it can appear via ports and consoles (such as JTAG, serial consoles or dedicated management ports) which are insufficiently secured. Second,  credentials of the administrator may be insufficiently protected. Another possibility is that the configuration module of the hardware and software might be insufficiently protected against malware injection and manipulation. 
 
 Physical attacks on the Open RAN deployment enables the retrieval of stored private keys, certificates, user plane data, control plane data and management data in cleartext. Moreover, attackers can try to modify the Open RAN components settings and configurations via local access in order to disable security features and allow eavesdropping or wiretapping on various planes, create performance issues \cite{5}.

Also for physical access, the required security standards should be developed. In addition, all stakeholders in the Open RAN system need to be identified, authenticated and trusted \cite{9}.
\end{itemize}


\section{Security Benefits of Open RAN} 
\label{sec:Benefits}








Besides all these risks, Open RAN brings of course a whole series of benefits. Several benefits are typical for Open RAN, others are also available in V-RAN and some of them are common for all 5G networks. Table \ref{benefits} provides an overview of the main benefits.

\begin{table*}[t]

\caption{Overview of Open RAN benefits}
\label{benefits}
\footnotesize
\begin{center}
\begin{tabular}{ |p{1.5cm} | p{4.5cm} | p{10cm}| }
\hline
\rowcolor{gray!30}
RAN type & Benefit & Short description \\
\hline
\hline
  Open RAN specific &	Full visibility	&The operator has increased visibility, allowing better security control and response to incidents. \cite{4,17,22}\\ \cline{2-3}
	&Selection of best modules&	Operators will be able to integrate best-in-class security platforms. \cite{4,13,17,22,41,5}\\  \cline{2-3}
	&Diversity&	Diversity and independency among the diverse modules will decrease the attack range. \cite{17,41,39}\\  \cline{2-3}
	&Modularity&	Enabling more efficient, seamless patch management and SW updates to remove vulnerabilities\cite{}. \cite{4,22}\\  \cline{2-3}
	&Enforcement of security controls&	security controls can be better enforced. \cite{13}\\ \cline{2-3}
	&Open interfaces&	Operators are independent of the supplier to react on security issues. \cite{7,41,17}\\ \cline{2-3}
	&Open source software&	Open source software has been verified by multiple parties. \cite{7}\\ \cline{2-3}
	&Automation	&The complete automation of network management can be speed up. \cite{4,7,41}\\ \cline{2-3}
	&Open standards	&Better coordination of security measures becomes possible. \cite{4,17}\\ \cline{2-3}
\hline
Also V-RAN&	Isolation&	Isolation enables more control and less issues during updates for security management. \cite{17}\\ \cline{2-3}
	&Increased scalabilty& 	It enables better trade-offs between performance and security. \cite{17}\\ \cline{2-3}
	&Control trust&	Operators are able to control full trust in their network. \cite{17}\\ \cline{2-3}
	&Less dependency between HW and network SW&	There are less risks for SW upgrades. \cite{22}\\ \cline{2-3}
	&Private networks&	There is easier migration to private networks. \cite{22,26}\\ \cline{2-3}
	&More secure storage of key material&	More secure storage of key material. \cite{22}\\
\hline
Also 5G&	Edge oriented&	Security is dealt closer to the edge of the network in order to stop attacks closer to the source. \cite{4}\\ \cline{2-3}
	&Simpler security model&	The zero trust security principle can be implemented. \cite{4}
   \\ 
   \hline
 \end{tabular}
\end{center}
\end{table*}

\subsection{Open RAN specific}

\subsubsection{Full visibility} Due to virtualization and the disaggregated components connected through open interfaces, operators have direct access to all network performance data and operational telemetry data representing activities between/within the different network functions. The integrity of this data is more ensured as this data is created isolated from the functions' executing environment. Combining this data with security log data results in an earlier detection of security problems and easier detection of the root cause \cite{4,17,22}. 

Note that full visibility can also be a risk. Due to the complexity, the root cause cannot always be easily detected and there is a danger that different vendors will not take accountability for potential issues. Following \cite{38}, it is claimed that the time and cost to perform a complete security review would seem to be multiplied by the number of vendors the operators take on board.

\subsubsection{Selection of best modules} 
Operators can more easily select the vendors offering the best products, meeting the required industry security standards and certifications \cite{4,17,22}. Examples of industry best practices are for instance ``secure by design'' DevSecOps in which information security operations are integrated into DevOps workflows and automated testing in development of containerized applications \cite{hsu2018hands}. The operator can also collaborate with the vendor to determine and influence Continuous Integration/Continuous Deployment (CI/CD) processes with continuous regression testing and software security auditing used by the supplier. Other good practices are the adoption of Supplier Relationship Management (SRM) with an inbound development process and strict security controls for Free and Open Source Software (FOSS), trust stack management with software coming from reliable supply chains and trusted, well-defined operations, intelligent vulnerability management, and multi-vendor System Integration (SI) with continuous verification on vendors sharing the same interpretation and implementation of functions \cite{4,13}.

There are a range of industry best practices that can be adopted including Groupe Speciale Mobile Association (GSMA), National Institute of Standards and Technology (NIST), European Union Agency for Cybersecurity (ENISA), National Telecommunications and Information Administration (NTIA), Center for Internet Security (CIS), Open Web Application Security Project (OWASP), Open Standards, Open Source (OASIS), national cyber security
organisations, Building Security In Maturity Model (BSIMM), Cloud Native Computing Foundation (CNCF), the Linux Foundation, SAFECode and CNTT \cite{17,22}.

\subsubsection{Diversity} Integrating independent and individual modules decreases the risk that common coding errors or practices of one single entity impact large parts of the network and thus decrease the attack range. Consequently, diversity helps to balance the security risks. Open RAN enables an expanded pool of vendors on the market, reducing a nation's dependence on any sole vendor for wireless services \cite{17,41,39}. \color{black} Despite the O-RAN entities being contrived following common and established standards, multiple vendors might embed diverse mechanisms, technologies to meet the standards or guarantees. This will eventually create competition among the vendors for better market returns. This competition can be considered healthy from the security perspective, as the standards might have to be improved from the security front to convince the consumers. \color{black}

\subsubsection{Modularity} Due to the modularity of the network, operators can switch to a CI/CD operating model, enabling seamless and effective patch management for fixing any detected security vulnerability. As a consequence, the vulnerabilities in the network are faster removed. In addition, updates become more transparent and have less impact on the overall network. Moreover, also operational agility is obtained making it possible to replace functional elements by new versions or capabilities \cite{4,22}. \color{black} The CI/CD method combined with the DevSecOps principles can insure individual modules carry out the updates or patches separately, eliminating any opportunity for complete compromise of the system in case of a malicious agent was conveyed via the updating process \cite{lee2021ran}. \color{black}

\subsubsection{Enforcement of security controls} Due to the choice among different vendors, modularity and open interfaces, the operator is in the position to demand strong security capabilities and control of its suppliers. For instance, in the case of a cloud architecture, the operator and the cloud infrastructure supplier have a common agreement in which this last one is responsible for the deployment of the latest security tools for detection and prevention \cite{13,17}.

\subsubsection{ Open interfaces} Open interfaces at the different levels give a higher exposure, resulting in more scrutiny and thus higher overall security. Thanks to the open interfaces, operators are not dependent anymore on the supplier in case of (security) issues and can do upgrades themselves, being able to react faster. It also gives the possibility to experiment with new functions and new vendors, exploring new ways to secure the network and its operation \cite{7,41,17}. 
This is at the same time a risk as in order to explore new possibilities by the operator, sufficient qualified people are required, which is not evident due to the complexity of the overall system.

\subsubsection{Open source software} Open-source software presents security challenges regarding its open nature but has the advantage of being verified by multiple independent parties, being rigorously and varied tested, and customized against threats \cite{7}. 

As mentioned before, the use of open source also includes many risks. It was concluded in the Github 2020 State of the Octoverse Report that vulnerabilities remain undetected in many cases for more than four years, before being disclosed \cite{bakhitova2020analysis}. Therefore, one cannot simply state that open source software is faster patched than proprietary software.

\subsubsection{Automation} The introduced intelligence in Open RAN can be used to automate the management and control via big data analysis, AI and ML. As a consequence, closed loop responses to changes in the network can be automatically performed. This has the advantage that no human interactions are required anymore, which inherently includes threats like humans accidentally altering the security posture of a network function or maliciously harvesting credentials, changing configurations, or implanting malware within the network \cite{4,7,41}.

Again, automation can bring risks as previously identified at the ML algorithms. 

\subsubsection{Open standards} Open RAN will be developed based on open standards, defined by the Open RAN consortium. Such standards enable to align on a common approach approved by leading members in the field and coordinate all information regarding security threats, vulnerabilities and exploits \cite{4,17}.

A prerequisite is of course the presence of these standards, which are not fully available at the moment. In addition, these standards should be correctly implemented.

\subsection{V-RAN specific}

\subsubsection{Isolation} Isolation is obtained via the defined interfaces between functional elements in an Open RAN. It offers on the one hand the possibility to insert controls for monitoring and on the other hand allows software updates and patches to be installed with less risk that version dependencies will create issues \cite{17,gavrilovska2020cloud}. 

\subsubsection{Increased scalability for security management} Often, there are trade-offs between application, performance and security requirements. Due to the modularity, operators can tailor their deployments and shift more easily the resources for monitoring and control to meet better to these requirements and improve scalability \cite{moreira2021task}. Also vRAN functional elements can be shifted to provide better isolation \cite{17}. 

\subsubsection{Control trust} Since operators control the platforms on which virtualized functions run in Open RAN, they have also complete control on the trust infrastructure \cite{benzaid2021trust}. The identity and provenance of each functional element is known and managed by using strong cryptographic mechanisms like signature operations. Each new version is validated by the operators and therefore they have control on what is running where on their networks \cite{17}.

However, it must be taken into account that the situation becomes more complex as there are more assets and stakeholders involved.

\subsubsection{Less dependency between HW and SW} In an Open RAN, there is less dependency between the network software and hardware. This makes it in the first place easier to perform the required upgrades in a faster way. Second, it also avoids risks associated to isolated security breaches \cite{22,gabilondo2022vnf}. 

\subsubsection{Private network} 
 Private 5G networks will soon become the general trend as they enable companies the possibility to fully customize the network according to their specific  needs with respect to speed, bandwidth, security requirements, on their locations and own timetable. It will enable companies to offer their customers a dedicated 5G experience, with applications in a large range of domains from healthcare, manufacturing, transportation, education, etc.
 Companies will have the option to build out and run their own private 5G network, or they can also outsource it to a mobile network operator or systems integrator \cite{siteM}. One such option is via network slicing, where each slice can be seen as a complete end-to-end network and includes the security capabilities according to the needs \cite{22}.
 
 There will be soon many players on the market to launch this innovative network as a service concept, replacing in many cases their existing Wi-Fi and fixed wireless/wired infrastructure.
 


\subsubsection{More secure storage of key material} In traditional network architectures, sensitive cryptographic key material such as for instance Access Stratum keys are more vulnerable to various threats as they are stored at the cell site \cite{cichonski5g}. In Open vRAN, this key material can be stored deep inside the network in a secure vCU, hosted in a data center \cite{22}.

\subsection{5G networks Related}
\subsubsection{ Edge oriented} Due to the open interfaces, the operator is able to spread the security analysis throughout the network and include monitoring at the edge. These edge-focused analytics will facilitate the detection and prevention of attacks at the lower part in the network in order to avoid DDoS and to block malicious data from reaching the core network. This is in particular important to support mobility services like services offered by IoT \cite{4, ge2017framework}.
\subsubsection{Simpler security model} In zero trust \cite{rose2020zero}, nothing is trusted unless it is verified, regardless of the location. The O-RAN Alliance completely embraces this principle. Therefore, everything needs to be verified and results need to be communicated \cite{4}. 
Zero trust networking can enhance security in different domains relying on robust standards. First it enables to  secure the technology and application stack including all interfaces and APIs. Second, it allows the leverage of the cloud-based nature of 5G and the deployment of cloud security functionality and telemetry. Third, it ensures the tailoring and customization of the security control via network slicing. Finally, it makes it possible to deploy multiple layers of authentication\cite{LIYANAGE2022103362}. 

\section{Lessons Learned and Discussion}
\label{sec:discussion}

\subsection{Lessons Learned}
In this section we summarize the key lessons learned for previous sections of this survey.

\subsubsection{Threat Vectors and Security Risks Associated with Open
RAN}
We have provided a clear taxonomy and an extensive overview of the different types of risks in Open RAN. There are basically three main domains of risks: process, technology and global. The global risks are general and apply for any type of RAN. In particular, we have shown that most of the technical risks follow from basic errors like insufficient mechanisms for encryption, authentication and authorization, improper configuration, software flaws, inappropriate event log management, lack of integrity and availability protection and unprotected physical access. Corresponding risk mitigation measures are provided, which are also related to the identified process risks. Basically, it all falls or stands with well defined standards and policies on all different processes covering the complete lifecycle, which can be clearly implemented, verified and audited in an automatic way. This is currently an ongoing work, but progress is on the way.

\subsubsection{Open RAN Best Security Practices}
As a derivative of C-RAN, Open RAN can inherit many security solutions and practices straight from C-RAN\cite{morais2020sdn, gavrilovska2020cloud}. However, due to its open architecture, it also requires a significant number of unique security solutions. Most of such solutions are required due to their lack of restrictions on O-RUs from different vendors. Open RAN enables blockchain based mutual authentication and privacy preserving P2P communication. Unlike conventional RAN technologies, Open RAN platforms can be upgraded with beamforming functionality and provide a countermeasure against eavesdroppers. Open RAN provides a platform to utilize the full benefit of AI algorithms for security solutions. In addition, many technology related attacks are caused by general design errors and can be mitigated by defining security standards and automation.   

\subsubsection{Security Benefits of Open RAN}
A significant amount of additional security benefits have been identified for Open RAN, compared to v-RAN and even 5G networks. However, most of them are closely related to the security risks. For instance, full visibility, selection of best modules, diversity, modularity and open interfaces also bring increased complexity and interdependency, requiring sufficiently trained people and trusted stakeholders. The same holds for open source software, which clearly has several advantages, but also brings several risks as identified here. Another example is the possibility to create an advanced level of automation in the network, but at the same time can lead to vulnerabilities from potential AI/ML attacks. Finally, for the enforcement of security controls and the adoption of open standards, the required standards, processes and policies still need to be fully defined.

\subsection{Discussion}
\color{black}
\subsubsection{Cost Of Security in O-RAN Deployments}

The RAN section of the telecommunication domain typically costs around 70\% to 80\% of the entire cost of the network; and it represent the best opportunity to reduce the cost of the network. In comparison, O-RAN is economically beneficial than PHY RAN or vRAN. The openness of the O-RAN is inviting the vendors to be more competitive with their apparatus, where 30\% less amount can be expected on Open Radios and O-RAN software. As proprietary BBUs are replaced by the typical servers, and their cost become less in comparison. According to \cite{fetterolf2021economic}, 32.5\% Capital Expenditure (CapEx) savings and 21\% Operational Expenditure (OpEx) savings can be expected regardless of whether the O-RAN is configured D or C setting. O-D-RAN in contrast to O-C-RAN has higher CapEx and OpEx values.

The major security repercussions of the O-RAN deployments are forecasted due to its openness of interfaces and modules. Ramifications for such flaws are typically automated security solutions or standards that define secure protocols for communication channels that cover confidentiality, integrity, and accountability; and an automated access control scheme. In addition, an AI or ML driven firewall and IDS facilities are imperative for securing the sub domains. Even with highest level of cryptographic primitives, computation power required for the secure communication protocols can be managed within the available O-RAN resources. The security overhead applicable to the channels are aggregated to the OpEx. The service oriented 5G networks are offering most of its services as cloud based services. Security can also be offered as a service where access control framework, firewall, and IDS facilities can be different flavors of the Security as a Service (SECaaS) use case \cite{ranaweera2020security}. These cloud oriented edge leveraged services eliminate the requirement for dedicated hardware; and nullify the required CapEx for launching SECaaS. Such a service can be purchased as a subscription based service that covers the entire O-RAN fronthaul domain, and it can be shared among the other O-RAN services. Hence, OpEx can be manageable. This service outsourcing will allow dynamic scaling ability. For security management and orchestration however, an agent of the SECaaS service should be deployed within the O-RAN for monitoring the security related actions and responses. Substantial amount of CapEx should be allocated for this agent as it requires to be high performing. Considering all these aspects where CapEx and OpEx are allocated for, the cost does justify the benefits granted through the SECaaS based services that would mitigate any disruptions to the O-RAN system.

\subsubsection{Impact of Quantum Computing}

\color{black} The Quantum Computing (QC) represents a superior computing power that exceeds almost 158 million times faster than a state-of-the-art supercomputer \cite{floridi2021digital}. The QC manifests an amalgamation of quantum mechanics of superposition, interference, and entanglement. The models that form the QC computations are often based on quantum bits (qbits), where the qbit value scopes higher than a typical computing bit, with its states. With this enormous computing power, QC is an obvious candidate for O-RAN processing core components, specifically for RIC-based computations. A QC enabled core RIC can deliver the real time outcomes of the Near-RT services. In spite of its power, QC based computing requires unique models for performing computing operations. Typical radio resource management and allocation computations might have to adhere to QC models or algorithms for solving the problems.

From the perspective of security, QC is a technology that challenges the complexity of modern applied cryptographic algorithms \cite{jurcut2020security}. The RSA algorithm, which is believed to be unbreakable in the current context, can be broken with a QC employing Shor's algorithm factoring discrete logarithms. Thus, QC poses a threat to the security of future networks and systems despite its rare existence. As a solution, Quantum Resistance (QR) cryptography was introduced and bares a significant interest among the research community. QR algorithms can be formulated from lattice-based, multivariate, hash-based, or elliptic curve based methods \cite{ranaweera2021survey}. Thus, O-RAN can leverage the capabilities offered through QC for achieving the guaranteed service level requirements. A Quantum Key Distribution (QKD) infrastructure can be adopted as the PKI for O-RAN internal entities including the xApps and rApps. On the contrary, QR based defense mechanisms should be adopted for signaling and critical channels for improving both the security and efficiency of the O-RAN.

\subsubsection{Role of Securing B5G/6G Applications}

Metaverse is a concept introduced to transcend physical space into a digital reality where all the possible actions in the real-world are enabled within the digitized world. Though the technologies Virtual Reality (VR), Augmented Reality (AR), Mixed Reality (MR), and Extended Reality (XR) are available, a holistic solution that incorporates all these digital visualization technologies is lacking \cite{lee2020unified}. Metaverse satisfies that void and envisages possibilities beyond measure. The network bandwidth, latency, jitter, availability, and reliability aspects of the RAN should be at its highest capacity to deliver a successful Metaverse \cite{lee2021all}. In addition, a significant level of interoperability and flexibility should be maintained within the network to guarantee performance with haptic sensory feedback. The O-RAN is capable of delivering the required flexibility and interoperability within its network. The security concerns of the Metaverse mostly exist in the virtual domain. Therefore, O-RAN cannot guarantee the internal security of the Metaverse; but can secure the network domain via typical security mechanisms.

Digital Twin (DT) technology contrives an exact replica of an object in the digital space. The replica or the twin, however, is formed in a simulated environment where all the actions committed by the actual object and its reactions can be mimicked on its digital counterpart \cite{jones2020characterising}. The main purpose of a DT application is to enable remote controlling and monitoring of apparatus or equipment situated in a factory environment align with Industrial IoT (IIoT) deployments \cite{ranaweera2022realizing}. The DT concept can be visualized as a miniature version of the Metaverse where an interactive interface is formed through AR-based collaborative tools. The O-RAN dynamic and flexible launching of xApps, with their Near-RT standards can facilitate the DT applications successfully. It requires lesser network based requirements than Metaverse. Though, the service rendered by the DT can be mostly critical, and require dedicated service channels with priority. Thus, fronthaul communication channels should embed better security credentials for DT deployments.

\color{black}

\section{Conclusion} 
\label{Sec:Conclusion}
In order to cope with the continuous growth of mobile subscribers, mobile data and mobile services, a drastically new approach is needed in order to ensure that the network resources are used in the most optimal way. In addition, this solution should particularly take into account a thorough security protection as also the amount and impact of cybersecurity attacks are continuously increasing. 

Open RAN offers all the possibilities to enable a great breakthrough in the network technology landscape and is able to address most of the current shortcomings in RANs thanks to the added openness and intelligence. However, due to this totally new approach where multiple vendors can now simultaneously integrate their technology, a more complex ecosystem exists resulting in a multitude of new risks and opportunities. We have provided a comprehensive overview of these different risks and benefits. We also discussed the best security practices to be applied.
As an important conclusion, in order to fully benefit from the most essential opportunities and to avoid the most important risks, the existence of an extended standard describing in detail the different processes in Open RAN is an essential step.

\section*{Acknowledgment}
This work is supported by 6Genesis Flagship (grant 318927) project. The  research  leading  to  these  results  partly  received  funding  from  European  Union's  Horizon  2020  research  and  innovation  programme  under grant  agreement  no   101021808 (H2020 SPATIAL project).  The  paper reflects only the authors' views. The Commission is not responsible for any use that may be made of the information it contains.


\bibliographystyle{elsarticle-num}
\bibliography{main}

\end{document}